%% file: main.tex
\documentclass{article}
\usepackage[utf8]{inputenc}
\usepackage{graphicx}
\usepackage{amssymb}
\usepackage{caption}
\usepackage{subcaption}
\usepackage{authblk}
\usepackage{rotating}
\usepackage{graphicx}
\usepackage{footnote}
\usepackage{xcolor}
\usepackage{soul}

\usepackage{url}
\usepackage{hyperref}
\usepackage{amsfonts}
\usepackage{amsmath}
\usepackage{placeins}

\usepackage[left]{lineno}
\usepackage{geometry}
\begin{document}

\title{Design and Implementation of the Cosmic Ray Tagger System for the ICARUS detector at FNAL}

\author[16]{A. Aduszkiewicz}
\author[3]{L. Bagby}
\author[4]{B. Behera\footnote{Now at South Dakota School of Mines}}
\author[1]{P. Bernardini}
\author[2]{S. Bertolucci}
\author[3]{M. Betancourt}
\author[12]{H. Budd}
\author[4]{T. Boone}
\author[14]{A. Campos}
\author[5]{D. Casazza}
\author[2]{V. Cicero}
\author[16]{D. Cherdack}
\author[15]{T.E. Coan}
\author[2]{L. Degli Esposti}
\author[2]{D. Di Ferdinando}
\author[5]{L. Di Noto}
\author[2]{C. Guandalini}
\author[2]{M. Guerzoni}
\author[4]{A. Heggestuen}
\author[4]{C. Hilgenberg\footnote{Now at University of Minnesota}}
\author[12]{R. Howell}
\author[6]{M. Iliescu}
\author[2]{G. Ingratta}
\author[3]{T. Kim}
\author[7]{U. Kose}
\author[2]{G. Laurenti}
\author[14]{C. Mariani}
\author[2]{N. Mauri}
\author[12]{K.S. McFarland}
\author[2]{E. Montagna}
\author[2]{A. Montanari}
\author[2]{N. Moggi}
\author[14]{M. Murphy}
\author[2]{L. Pasqualini}
\author[13]{V. Paolone}
\author[2]{L. Patrizii}
\author[2]{G. Pellegrini}
\author[2]{V. Pia}
\author[2]{F. Poppi}
\author[2]{M. Pozzato}
\author[2]{A. Ruggeri}
\author[8]{P. Sala}
\author[9]{P. Sapienza}
\author[3]{A. Schukraft}
\author[2]{G. Sirri}
\author[10]{L. Stanco}
\author[7]{A. Surdo}
\author[2]{M. Tenti}
\author[2]{V. Togo}
\author[2]{N. Tosi}
\author[11]{M. Vicenzi}
\author[4]{R. J. Wilson}
\author[2]{S. Zucchelli}

\affil[1]{INFN Sezione di Lecce and University of Salento, Lecce, Italy}
\affil[2]{INFN Sezione di Bologna and University of Bologna, Bologna, Italy}
\affil[3]{Fermi National Accelerator Laboratory, Batavia, IL 60510, USA}
\affil[4]{Colorado State University, Fort Collins, CO 80523, USA}
\affil[5]{INFN Sezione di Genova and University of Genova, Genova, Italy}
\affil[6]{INFN LNF, Frascati, Italy}
\affil[7]{ETH Zurich, Switzerland}
\affil[8]{INFN Sezione di Milano, Milano, Italy}
\affil[9]{INFN LNS, Catania, Italy}
\affil[10]{INFN Sezione di Padova and University of Padova, Padova, Italy}
\affil[11]{Brookhaven National Laboratory, Upton, NY 11973, USA}
\affil[12]{University of Rochester, Rochester, NY 14627, USA}
\affil[13]{University of Pittsburgh, Pittsburgh, PA 15260, USA}
\affil[14]{Center for Neutrino Physics, Virginia Tech, Blacksburg, VA 24061, USA}
\affil[15]{Southern Methodist University, Dallas, TX 75275, USA}
\affil[16]{University of Houston, Houston, TX 77004, USA}
    


 
 
 
 



\maketitle

\abstract{The ICARUS-T600 Liquid Argon Time Projection Chamber is operating at Fermilab at shallow depth and thus exposed to a high flux of cosmic rays that can fake neutrino interactions. A cosmic ray tagging (CRT) system ($\sim$1100 m$^2$), surrounding the cryostat with two layers of fiber embedded plastic scintillators, was developed to mitigate the cosmic ray induced background. Using nanosecond-level timing information, the CRT can distinguish incoming cosmic rays from outgoing particles from neutrino interactions in the TPC.  In this paper an overview of the CRT system, its installation and commissioning at Fermilab, and its performance are discussed.}


\section{Introduction }
\input{Introduction_preliminary}

\section{The ICARUS Cosmic Ray Tagging System }\label{sec:overviewCRT}

\input{overview_preliminary}

\subsection{The Top CRT}
\input{TopCRTDesignNew}\label{sec:TopCRT}

\subsection{The Side CRT}
\input{SideCRTDesign}

\subsection{The Bottom CRT}
\input{BottomCRT}

\section{Installation and Commissioning}\label{sec:instandcomm}
\input{CommissioningIntroduction}

\subsection{Installation of Overburden}
\input{OBinstallation}

\subsection{Timing Calibration}\label{sec:calibT}
\input{TimingCalibration}

\subsection{Detector Response Calibration} 
\input{DetectorResponseCalibration}\label{sec:calibR}

\section{CRT Hit Reconstruction}\label{sec:reco} 
\input{CRTHitReconstruction.tex}


\section{CRT Performance}\label{sec:results}

\subsection{Time Resolution}
\input{timeResolution.tex}

\subsection{Module Efficiency}
\input{efficiency.tex}

\subsection{System Tagging Efficiency}
\input{tagging.tex}




\section{Conclusions}
\input{Conclusions.tex}
\section{Acknowledgments}
This document was prepared by the ICARUS collaboration using the resources of the Fermi National Accelerator Laboratory (Fermilab), a U.S. Department of Energy, Office of Science, HEP User Facility. Fermilab is managed by Fermi Research Alliance, LLC (FRA), acting under Contract No. DE-AC02-07CH11359.
This work was supported by the US Department of Energy, INFN, EU Horizon 2020 Research and Innovation Program under the Marie Sklodowska-Curie Grant Agreement No. 734303, 822185, 858199, and 101003460 and Horizon Europe Program research and innovation programme under the Marie Sklodowska-Curie Grant Agreement No. 101081478. Part of the work resulted from the implementation of the research Project No. 2019/33/N/ST2/02874 funded by the National Science Centre, Poland. The ICARUS Collaboration would like to thank the MINOS Collaboration for having provided the side CRT panels as well as Double Chooz (University of Chicago) for the bottom CRT panels. 
We also acknowledge the contribution of many SBND colleagues, in particular for the development of a number of simulation, reconstruction, and analysis tools which are shared within the SBN program.

\input{references}
\bibliographystyle{unsrt}

\end{document}

%% file: Introduction_preliminary.tex
The Short Baseline Neutrino (SBN) program employs Liquid Argon Time Projection Chambers (LArTPC) to sample neutrinos along the Fermilab Booster Neutrino Beam (BNB). Its goal is to assess the possible existence of sterile neutrino states. This investigation involves measuring the $\nu_e$ appearance and the $\nu_{\mu}$ disappearance oscillation channels using the Short Baseline Near Detector (SBND) and the ICARUS-T600 Far Detector at 110 m and 600 m from the neutrino source, respectively \cite{sbnproposal}. Additionally, the ICARUS detector is exposed 6$^\circ$ off-axis to neutrinos from the Main Injector (NuMI) beam. The ICARUS-T600 detector consists of two identical cryostats filled with 760 tons of liquid argon. Each cryostat is divided into two TPCs with a common central cathode at 75 kV producing an electric field of 500 V/cm along a 1.5 m drift length. Charged particles interacting in LAr produce ionization charge and scintillation photons. To readout the ionization charge each TPC has three parallel anode wire planes at different orientations (0$^{\circ}$,~$\pm$~60$^{\circ}$) with respect to the horizontal \cite{ICARUS}. By combining the wire coordinates at the same drift time a 3D track reconstruction with a O(1 mm) resolution is achieved. Additionally, 360 photomultiplier tubes (PMTs) deployed in groups of 90 devices each are installed behind the wire planes to detect the scintillation light produced by charged particles
in LAr and are used for  triggering  the detector \cite{PMTs}.
\par The ICARUS-T600 was operated for the first time underground at the INFN Gran Sasso National Laboratory, where the rock overburden (3700 m water-equivalent) reduces the cosmic ray muon flux  by 10$^6$  with respect  to the sea level flux. At Fermilab the detector is located on the surface. High energy photons generated by cosmic muons interacting with the surrounding materials imposes a background for identifying $\nu_e$ candidates, since electrons produced via Compton scattering/pair production can mimic signal a $\nu_e$ CC event signal. In order to mitigate the cosmogenic induced background, a concrete overburden (6 m.w.e.) is installed on top of ICARUS. The cosmic ray flux through the ICARUS TPC with and without the overburden was evaluated by a Monte Carlo simulation \cite{Behera:2021iap}. Table \ref{tab:obreduction} reports  the expected rates of cosmic particles crossing the active LAr for kinetic energy $>$50 MeV \cite{poppiPhd}.
The overburden reduces the dominant muon flux by $\sim$~25$\%$, stopping muons with a kinetic energy ~$E_K\leq$1.5~GeV. The suppression is more effective for hadrons, with a reduction by a factor~$\sim$~200 for neutrons and $\sim$~500 for protons. The gamma rays are almost fully suppressed.
About  11  cosmic muons cross the detector during the $\sim$~1 ms drift time. A synchronization between the cosmic ray tagger and the ICARUS photon detection system with a few nanosecond accuracy is therefore required to tag cosmic particles recorded during the beam spill.

\begin{table}[b]
\centering
\begin{tabular}{|cccc|}
\hline
\textbf{Particle} & \textbf{w/o OB [Hz]} & \textbf{w/ OB [Hz]} & \textbf{reduction factor without/with OB} \\ \hline
$\mu^\pm$         & 17117                    & 12761                 & 1.34                    \\
p                 & 54                   & 0.10              & $>$ 500                 \\
$\gamma$          & 116                   & 0.03                    & $>$ 3500                      \\
n                 & 1426                    & 6.8                & $>$ 200                  \\
\hline
\end{tabular}
\caption{Estimated rate of cosmic particles at ground level with E$_k$$>$50 MeV entering the LAr active volume without (w/o)  and with  (w/) the concrete overburden (OB).}
\label{tab:obreduction}
\end{table}

 This paper is presented as follow: In Section \ref{sec:overviewCRT}  the ICARUS cosmic ray tagging system is described.  The installation and commissioning are presented in Section \ref{sec:instandcomm}. In Sections \ref{sec:calibT} and \ref{sec:calibR}  the timing and detector response calibration are reported. The CRT hit reconstruction is described in Section \ref{sec:reco} while the time resolution and efficiency are reported in Section \ref{sec:results}.

%% file: overview_preliminary.tex
\begin{figure}
    \centering
    \includegraphics[width=\textwidth]{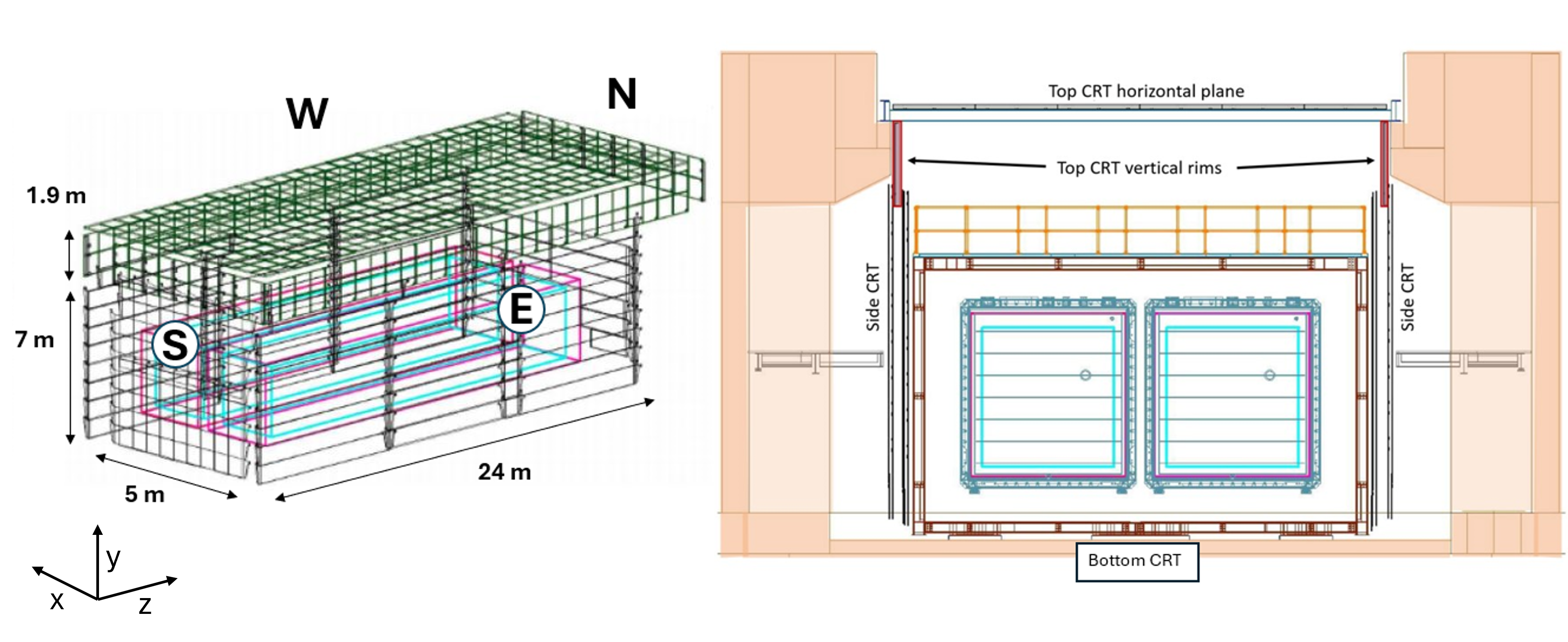}
    \caption{(Left) CRT layout including the cold vessels (magenta) and the argon active volumes (cyan). The Top (green) and the Side (black) CRT subsystems are also sketched. (Right) A view of the CRT system surrounding the ICARUS detector in a plane transverse with respect to the neutrino beam.}
    \label{fig:CRTlayout}
\end{figure}

The cosmic ray tagging (CRT) system is divided into three different sub-systems - Top, Side, and Bottom CRT - each deployed over different regions of the TPC. In Figure \ref{fig:CRTlayout} a drawing of the CRT system with the Top and Side CRT sub-systems is shown. The three  sub-systems complement each other and are comparable in terms of timing and spatial resolution. 

%% file: TopCRTDesignNew.tex
\begin{figure}
    \centering
    \includegraphics[width=0.6\textwidth]{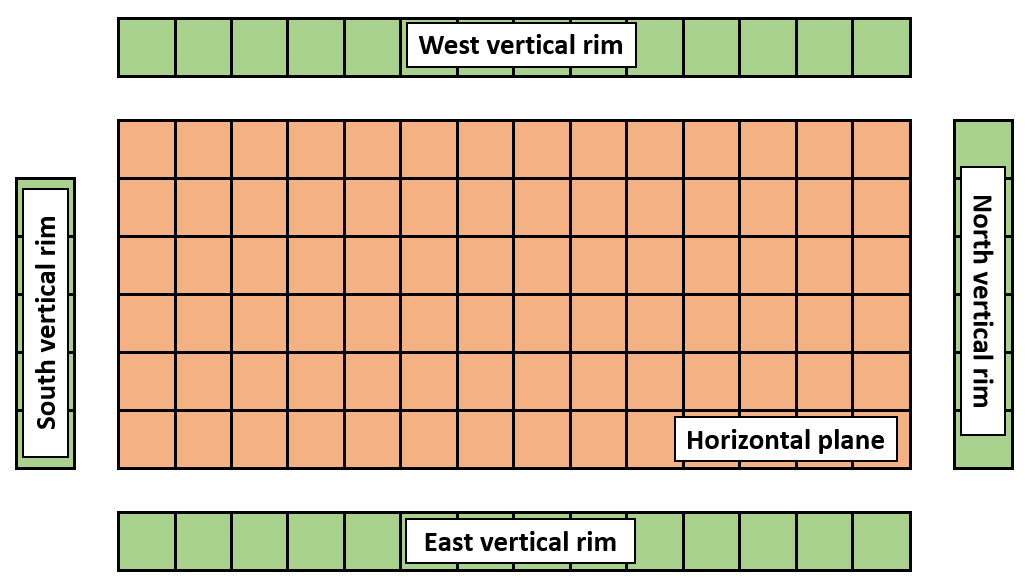}
    \caption{Layout of the Top CRT modules. Orange modules are arranged horizontally on top of the cryostat, green modules are installed vertically.}
    \label{fig:TopCRTregion}
\end{figure}
\begin{figure}
    \centering
    \includegraphics[width=.7\textwidth]{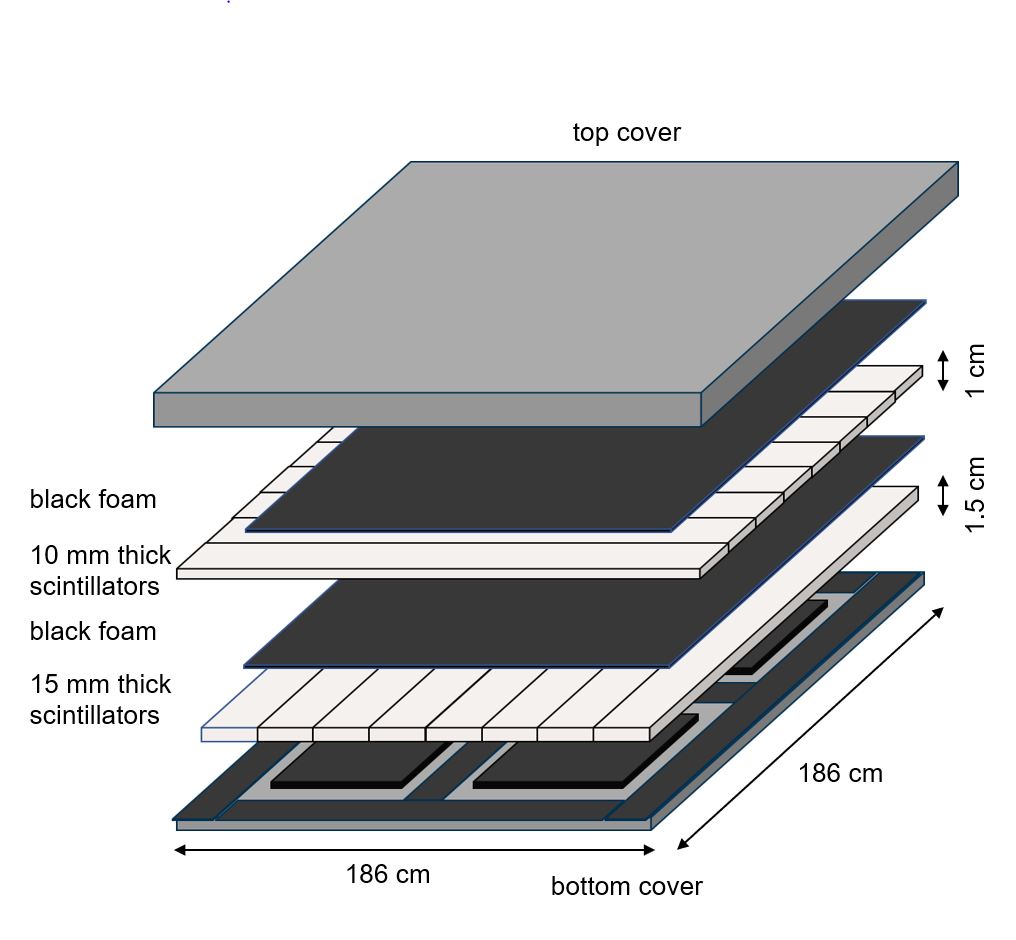}
    \caption{Design of a Top CRT module.}
    \label{fig:TopCRTsketch}
\end{figure}

The Top CRT is made of 123 modules with a surface area of 426 m$^2$: 84 modules are installed on the top horizontal plane, while 39 modules cover the upper area of the TPC (vertical rims). In Figure \ref{fig:TopCRTregion} the layout of the installed Top CRT is shown. Each module is a hodoscope consisting of two orthogonal layers of eight scintillator bars all encased in 1.86~m $\times$ 1.86~m aluminum box. The scintillator bars are 23~cm wide, 1.84~m long and are 10 and 15~mm thick for the outer and inner layers, respectively. There are 1968 bars in the Top CRT system. An illustration of a Top CRT module is provided in Figure \ref{fig:TopCRTsketch}. 

\begin{figure}
    \centering
    \includegraphics[width=.6\textwidth]{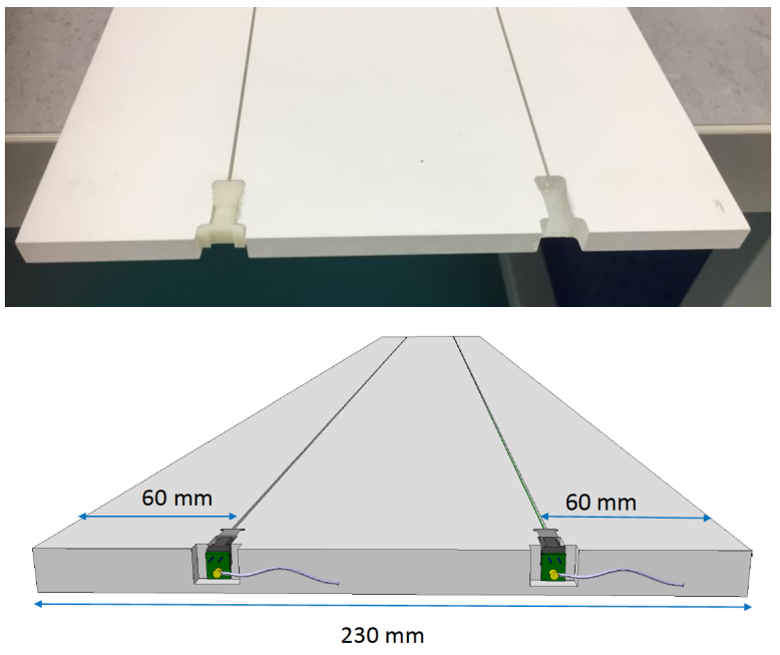}
    \caption{Top: picture of the 10 mm thick scintillators, with a detail of a SiPM connector on the left and an empty socket on the right. Bottom: illustration of the fiber positions along the scintillator bar.}
    \label{fig:ScinGrooves}
\end{figure}
Each scintillator strip is instrumented with two WaveLength Shifter (WLS) Kuraray Y-11 fibers \cite{WLSfibre} with a diameter of 1 mm embedded along the longitudinal direction of the bar and located 6 cm from each side, as sketched in the illustration in Figure \ref{fig:ScinGrooves} (bottom). 
The fibers are readout from one end, the opposite end-side is mirrored in order to enhance the light yield. The mirroring is obtained by polishing the end of the fiber and then coating it with aluminum reflective layers a few micron thick by aluminum sputtering in vacuum. The WLS fibers were glued in the scintillator groove with an epoxy-based hard optical cement \cite{EJglue}. In total $\sim$~7400~meters of fibers were used for the construction of the Top CRT system.
\begin{figure}
    \centering
    \includegraphics[width=0.7\textwidth]{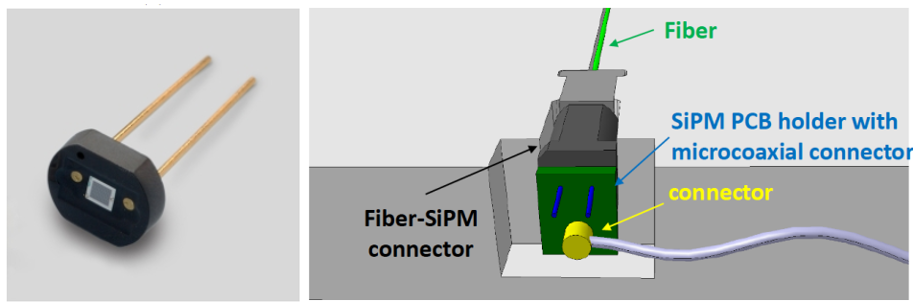}
    \caption{Left: Picture of a Hamamatsu S13360-1350CS SiPM. Right: illustration of the SiPM connection scheme to WLS fiber.}
    \label{fig:SiPM}
\end{figure}
The light read-out is performed by Hamamatsu S13360-1350CS silicon photomultipliers (SiPMs) (Figure \ref{fig:SiPM} left), with an active area of 1.3~mm~$\times$~1.3~mm \cite{sipm}.
The SiPMs were plugged and glued using a soft silicon-based optical glue \cite{Bluesil} into a custom-made SiPM-holder and connected to the scintillator bar and to the WLS fiber (Figure \ref{fig:SiPM} right). SiPMs pins are connected to a printed circuit board (PCB) and routed via 50 $\Omega$ micro-coaxial cables 1.8~m long, carrying both signal and bias voltage to a PCB patch panel connected to the readout electronics. Overall, 4000 SiPMs were used. The inner surface of the aluminum box was covered with black sponge-like tape and each scintillator layer was separated by a black foam-like plastic sheet. The assembled scintillator strips were wrapped with Tedlar$^{\circledR}$ strips to prevent reflected light from reaching the scintillators and the SiPMs, thus ensuring light tightness. In order to avoid any shift of the scintillator bars, plexiglass spacers 3 mm thick were positioned between the scintillator edge and the aluminum box lateral sides.
\begin{figure}
    \centering
    \begin{subfigure}[t!]{.8\textwidth}
        \includegraphics[width=.9\textwidth]{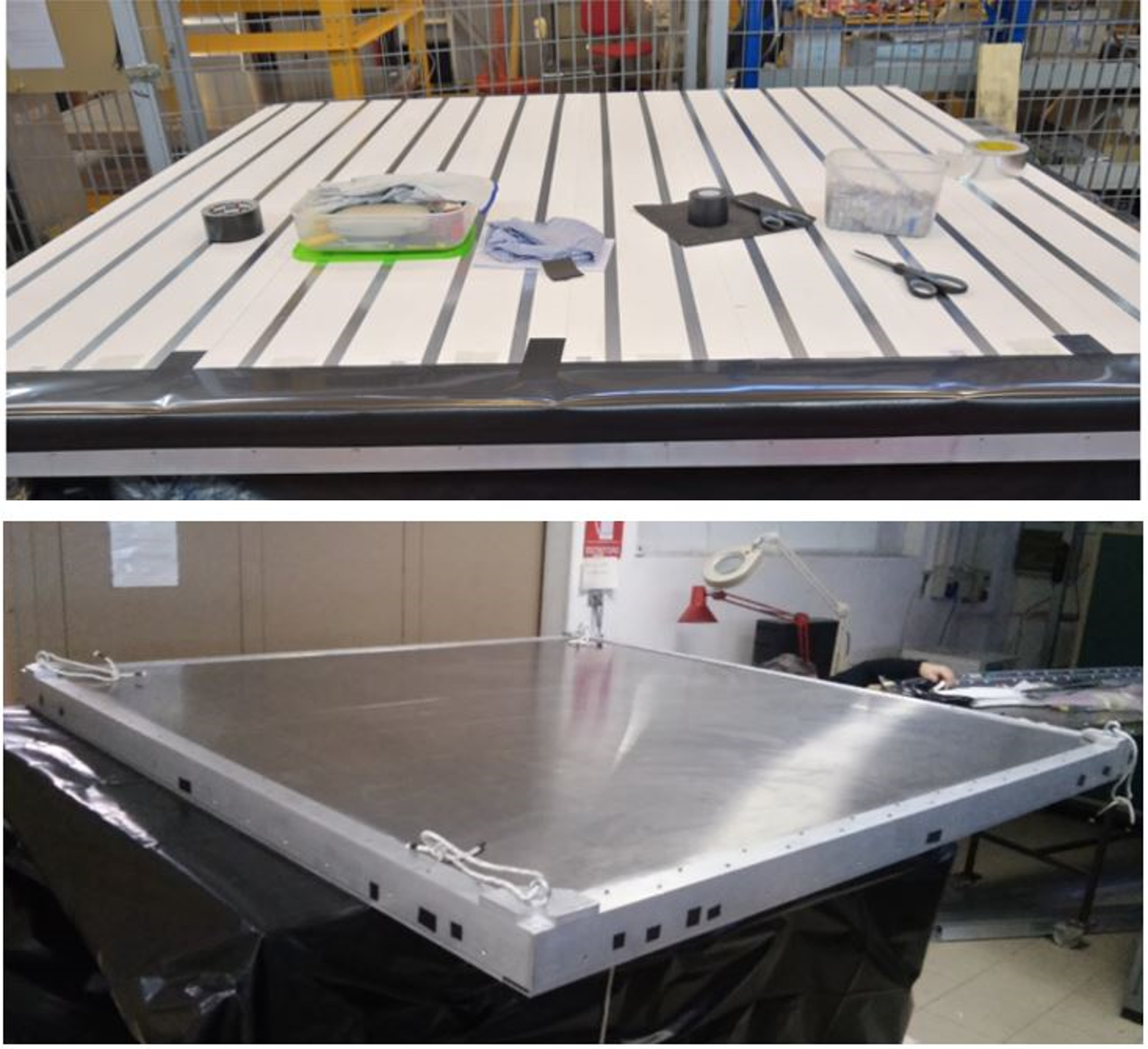}
    \end{subfigure}
    \caption{Different assembly stages of the CRT module: (top) picture of the assembled scintillator hodoscope of the Top CRT; (bottom) picture of the module inside its aluminum case.}
    \label{fig:wrapping}
\end{figure}
The analog signals of 32 SiPMs from each module are acquired by a CAEN DT5702 Front End Board (FEB) \cite{feb}. The board provides two LEMO connectors, T0 and T1, for timing reference signals. On each bar, a coincidence between the signals of the couple of SiPMs is generated with a threshold of 1.5 photoelectrons (p.e.). A trigger is issued by a coincidence logic between the two module layers. At the occurrence of a trigger, all the 32 channels of an  FEB are digitized and read out, and two event timestamps (T0 and T1) are generated. During the digitization and storing of the signal pulse heights, a dead time of 22 $\mu$s is generated. The board communicates with the host computer through ethernet protocol, which allows connecting multiple boards with each other in daisy chain mode.

The assembling and testing of the modules were performed at the INFN laboratories in Frascati, Italy. The construction started in April 2019 and was concluded in February 2020. In Figure \ref{fig:wrapping} pictures of different assembly stages of the CRT modules are shown. 

Two extra CRT modules 15 cm apart were installed above the concrete overburden, to create an external hodoscope. This hodoscope
is mobile, facilitating measurements at different detector location for calibration and efficiency studies. 


%% file: SideCRTDesign.tex
\begin{figure}
    \centering
    \includegraphics[width=.6\textwidth]{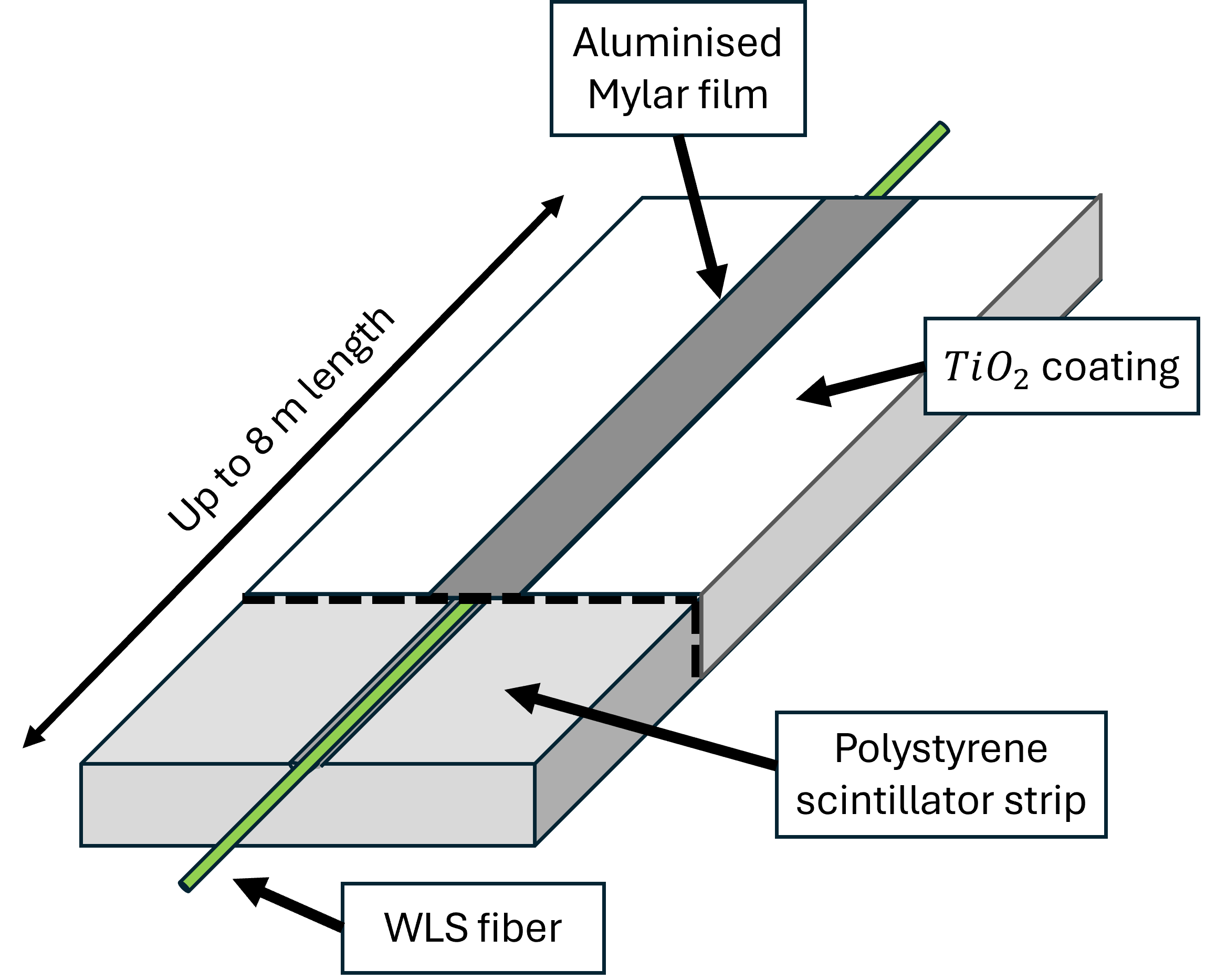}
    \caption{Schematic of a MINOS scintillator strip from \cite{MINOS}.} 
    \label{fig:MINOSstrip}
\end{figure}
\begin{figure}
  \centering
   \includegraphics[width=0.95\textwidth]{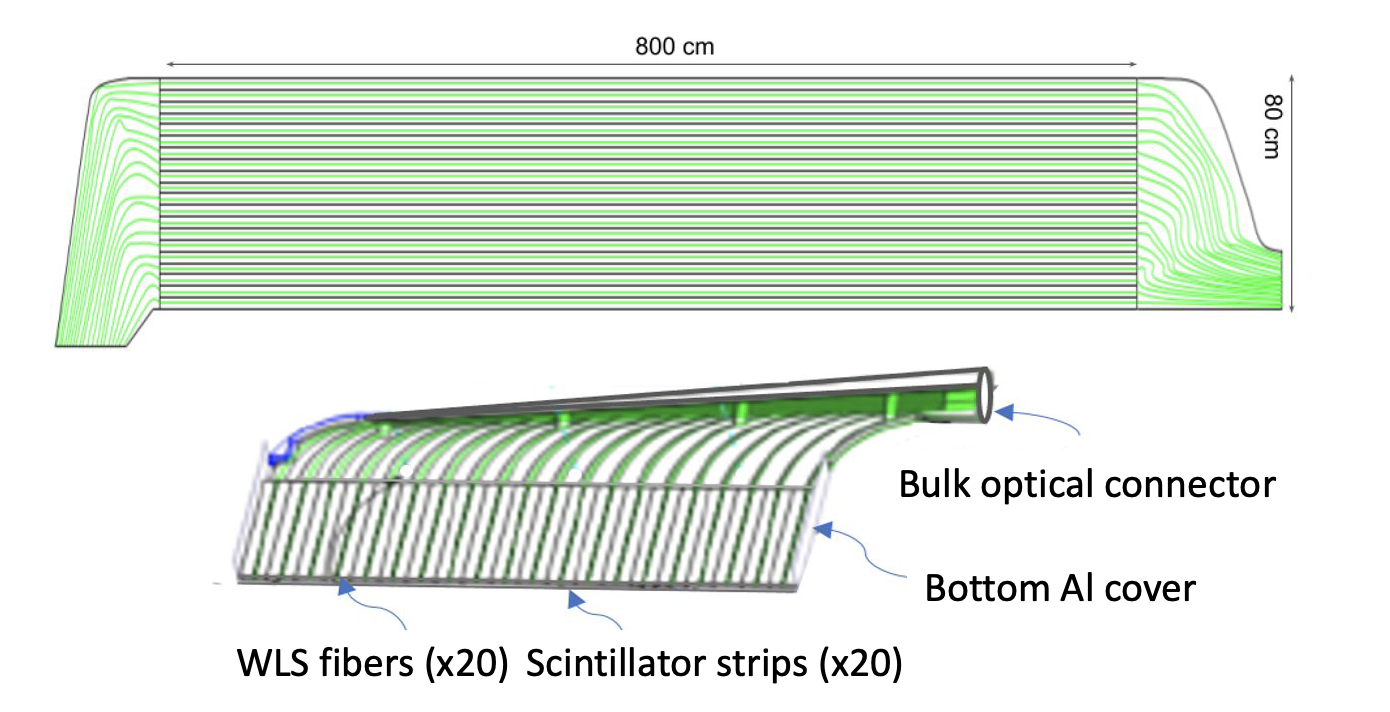}
   \caption{ MINOS scintillator module schematic used in the Side CRT system with a detailed view of WaveLength Shifter fiber-embedded scintillator  }. 
   \label{fig:MINOSmodule}
\end{figure}
\begin{figure}
   \centering
    \includegraphics[width=0.60\textwidth]{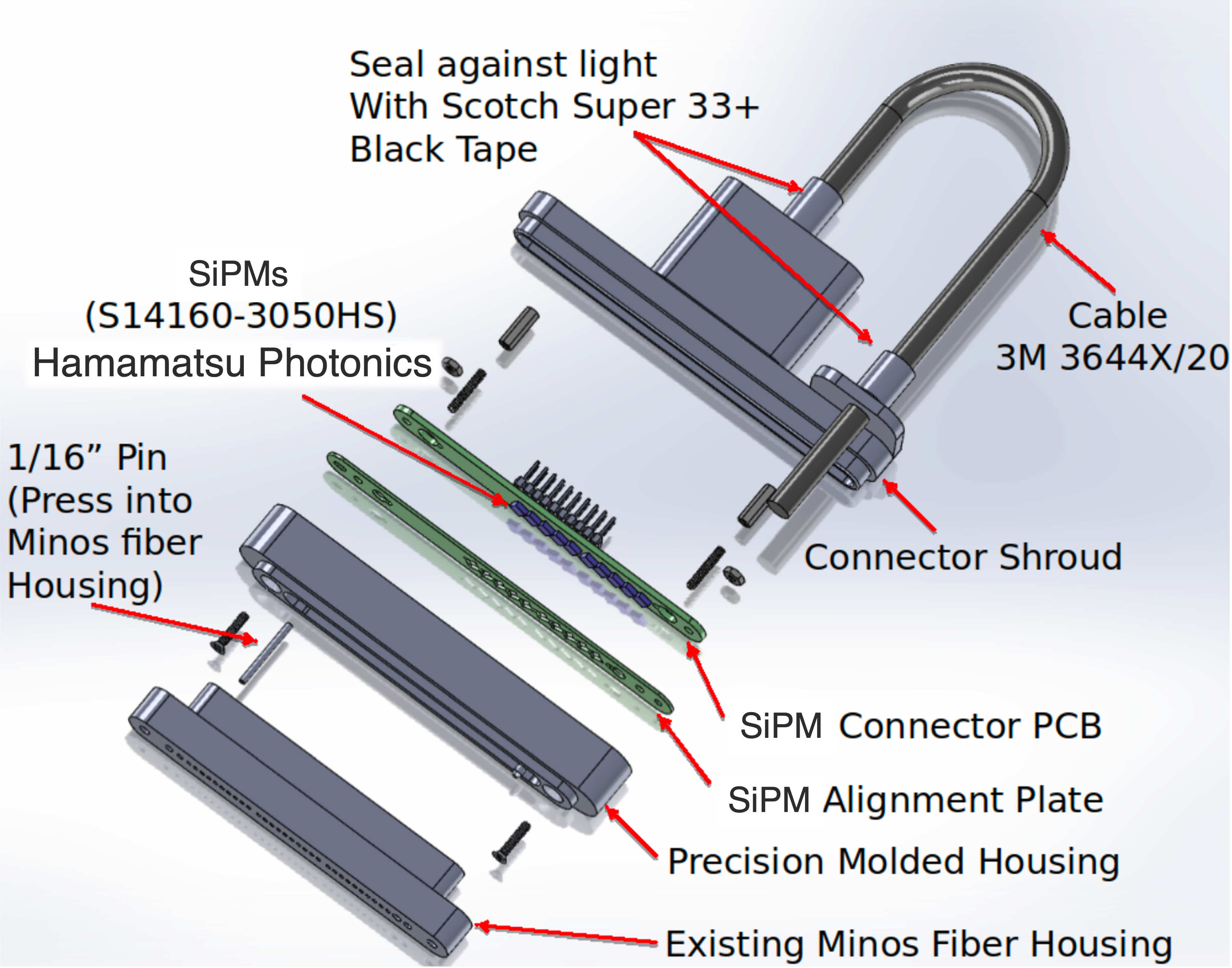}
    \caption{ An expanded drawing of all components in a Side CRT Optical Readout Module (ORM) from C. Hilgenberg.}
    \label{fig:ORM}
\end{figure}

The Side CRT utilizes repurposed scintillator modules from the decommissioned MINOS experiment \cite{MINOS}. The modules are 8~m long, 80~cm wide and 1~cm thick; each containing 20 scintillator strips 4 cm wide. Figure \ref{fig:MINOSmodule} shows a schematic of the module and the WaveLength Shifteer fiber-embedded scintillator . 
Figure \ref{fig:MINOSstrip} shows a schematic of  a MINOS scintillator strip used for the Side CRT modules.
The wave-length shifting fibers of one module are grouped together and coupled to a Hamamatsu S14160-3050HS Multi-Pixel SiPM by an optical connector (see Figure \ref{fig:ORM}).
An array of 10 SiPMs embedded in the plastic housing are referred to as an Optical Readout Module (ORM) and reads out 20 fibers in adjacent pairs. Fibers are then readout on both ends of the scintillator strip, apart from a small number of cut MINOS modules that readout at a single end with a reflector at the opposite end. With each SiPM reading out a pair of fibers, the Side CRT has a total of 2,710 readout channels across 93 FEBs. 

The CAEN DT5702 FEB, the same as for the Top CRT, digitize signals from one end of up to 3 modules. Due to length of the modules, signals from the other end of a module are read by another FEB. Coincidence between two layers of scintillator modules is enabled by an external network between the FEBs.
\par
\begin{figure}
    \centering
    \includegraphics[width=0.80\textwidth]{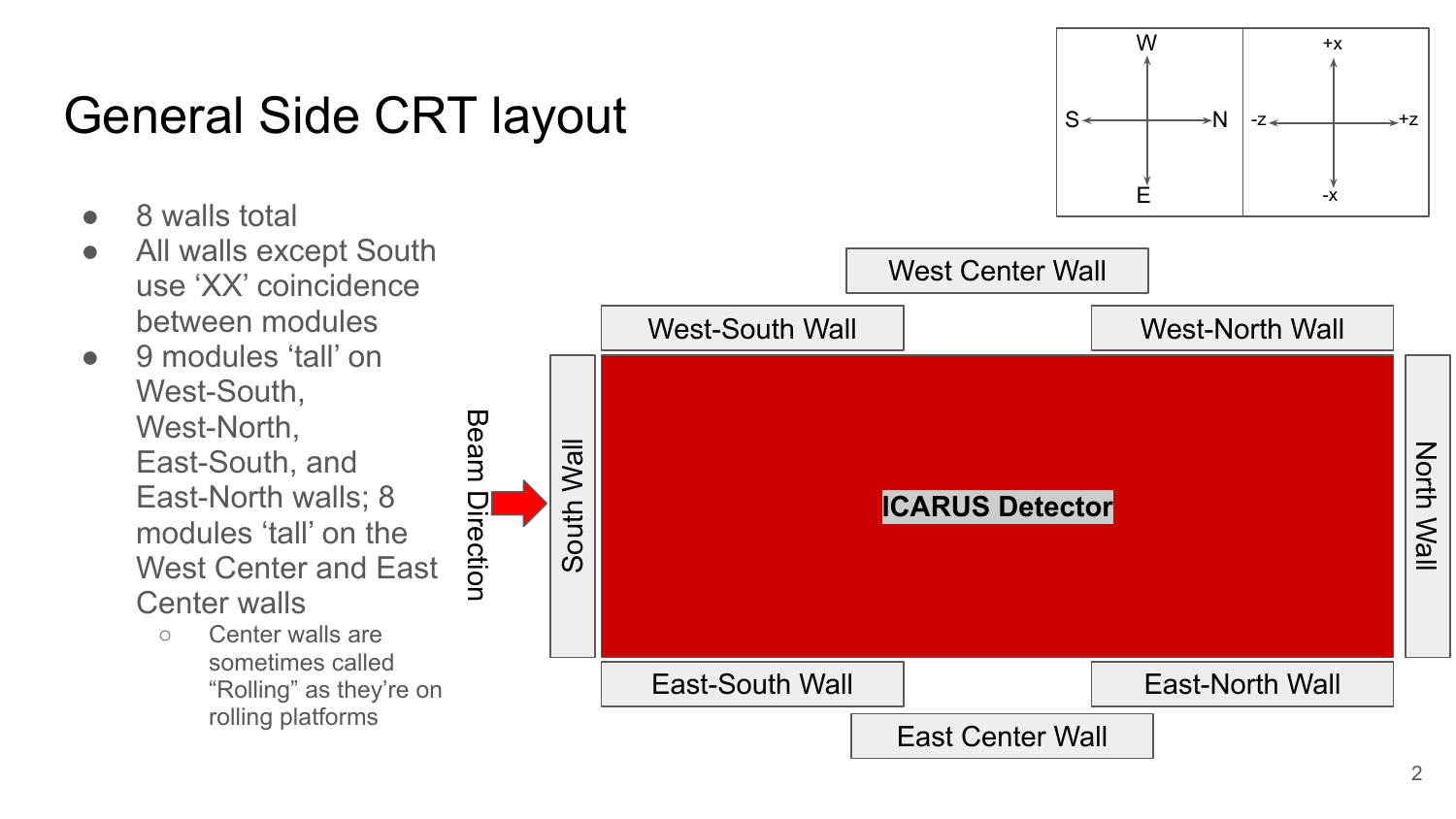}
    \caption{Overview of the eight walls composing the side CRT system.} 
    \label{fig:sideCRTgeometry}
\end{figure}

 The Side CRT consists of eight `walls': North, South, West-North, West-Center (Rolling), West-South, East-North, East-Center (Rolling), and East-South; Figure \ref{fig:sideCRTgeometry} shows a schematic of the Side CRT with the different walls. 
Each wall is made up of an internal and an external module operating in coincidence. Except for the South wall, the overlapping layers are arranged to form an `XX coincidence' with the scintillator strips in the internal and external layers parallel to each other. The South Wall has the outer layer placed vertically to form an `XY coincidence' with the inner layer. \par

The East and West Side CRT regions are further divided into East/West-North, East/West-Center, and East/West-South; the first direction indicates whether the wall is on the east or west side of the TPC, then the second direction gives the relative position of that wall along the north/south axis. The center walls (East and West) are both on rolling platforms and sit slightly outward  of the other sections on a given side. The rolling platforms allow access to the readout electronics and module ends which would otherwise sit behind the center wall. Due to overhanging cable trays which run from off-detector equipment to the top of the detector, the Center walls have 8 modules along the height, while the other walls on the East/West side are composed by 9~modules. On all walls with XX coincidence, a $\sim$~2.5 cm vertical offset was added between layers to avoid having gaps between strips in the modules line up with each other. 



%% file: BottomCRT.tex
\label{BottomCRT}
\begin{figure}[!t]
    \centering
    \includegraphics[width=0.7\linewidth]{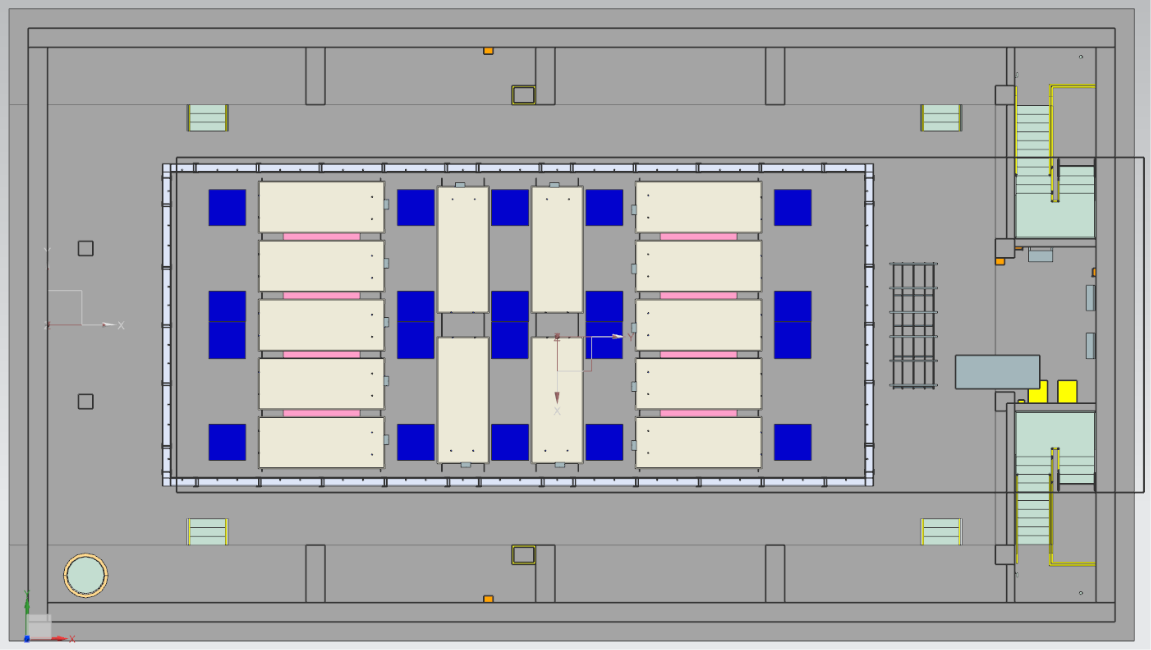}
    \caption{Positioning of Bottom CRT modules (light yellow).}
    \label{fig:bottomLayout}
\end{figure}
\begin{figure}[!t]
    \centering
    \includegraphics[width=0.9\linewidth]{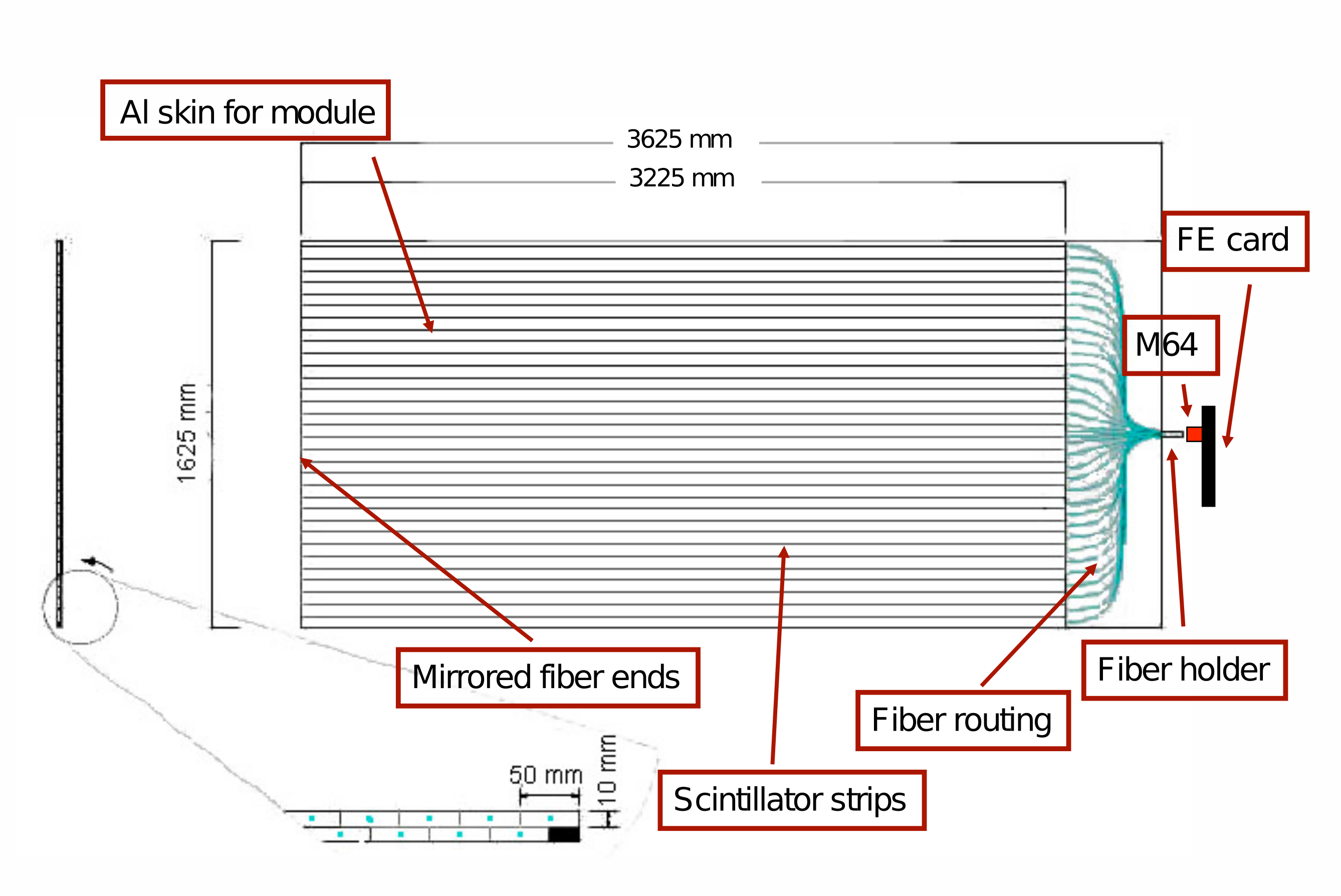}
    \caption{Schematic of a Bottom CRT module from Ref.~\cite{DCpaper}. The main drawing shows the 32 scintillator strips making up one layer of the module, the fibers that run through them, the MAPMT they are connected to, and the FEB. The inset at the bottom shows how the individual strips are placed in two layers, which are offset from each other.}
    \label{fig:bottomModule}
\end{figure}
To implement the Bottom CRT, nine modules built for the Double Chooz outer veto detector~\cite{DCpaper} were installed below the ICARUS cryostat, as sketched in Figure~\ref{fig:bottomLayout}.

Each module consists of 64 strips of polystyrene plastic scintillator (3.2 m long, with a cross section of 50 mm $\times$10 mm), arranged in two staggered layers as shown in Figure~\ref{fig:bottomModule}. Each strip is equipped with a wavelength-shifting fiber (Kuraray Y-11 multi-clad non-S type) running through a hole extruded at the center of the strip along its length. Each strip has a reflective coating of 0.25 mm of TiO$_2$, and the fibers, read at one end only, are mirrored at the other end to increase the light yield. Each of the 64 fibers of a single module are individually attached to a single channel of an 8$\times$8 Hamamatsu H8804 multi$-$anode photomultiplier tube (MAPMT). Each MAPMT, with 64 individual channels, is powered by a single HV line. 

\begin{figure}[!t]
    \centering
    \includegraphics[width=0.9\linewidth]{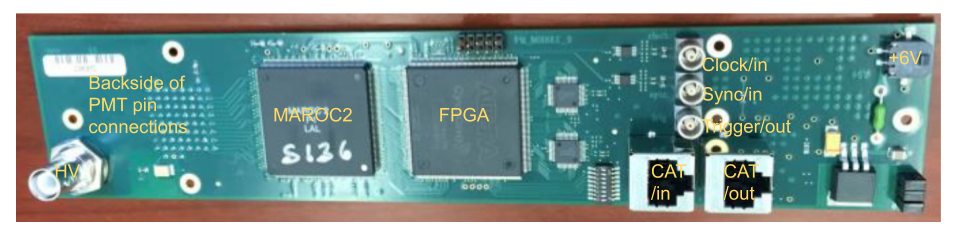}
    \caption{Picture of and FEB module motherboard showing the MAROC2 chip, FPGA, connections for clock, sync, trigger, and the CAT6 daisy-chain. The HV connector for the MAPMT is on the left, and the 6~V connection to power the FEB is on the right.}
    \label{fig:bottomFEB}
\end{figure}

Each MAPMT is directly connected to a FEB, as shown in Figure~\ref{fig:bottomFEB}, comprising several components. It includes a MAROC2 ASIC~\cite{MAROC}, enabling individual amplification adjustment for each channel. The FEB is controlled by an FPGA, which also acts as a buffer for the data and records hit timestamps~\cite{DCpaper}. The timestamp is determined by a 32-bit counter, which is incremented every 16~ns by a 62.5~MHz clock signal. This counter is reset every 7 seconds based on a external pulse-per-second (PPS) signal, described in section \ref{sec:calibT}, which acts as an absolute time reference. At a later stage the DAQ combines the timing information to create a full 64-bit timestamp for each hit with a 16~ns resolution. The ADC and timestamp data of each hit is then sent along a Ethernet daisy-chain to a back-end custom USB module, which is connected to a computer. The custom USB module is equipped with an FPGA and it is responsible for processing the data, adding an independent timestamp to the data stream, converting the ethernet signal, sending and receiving commands to the FEBs, and passing the data to the computer.

%% file: CommissioningIntroduction.tex
\begin{figure}
    \centering
    \includegraphics[width=.5\textwidth]{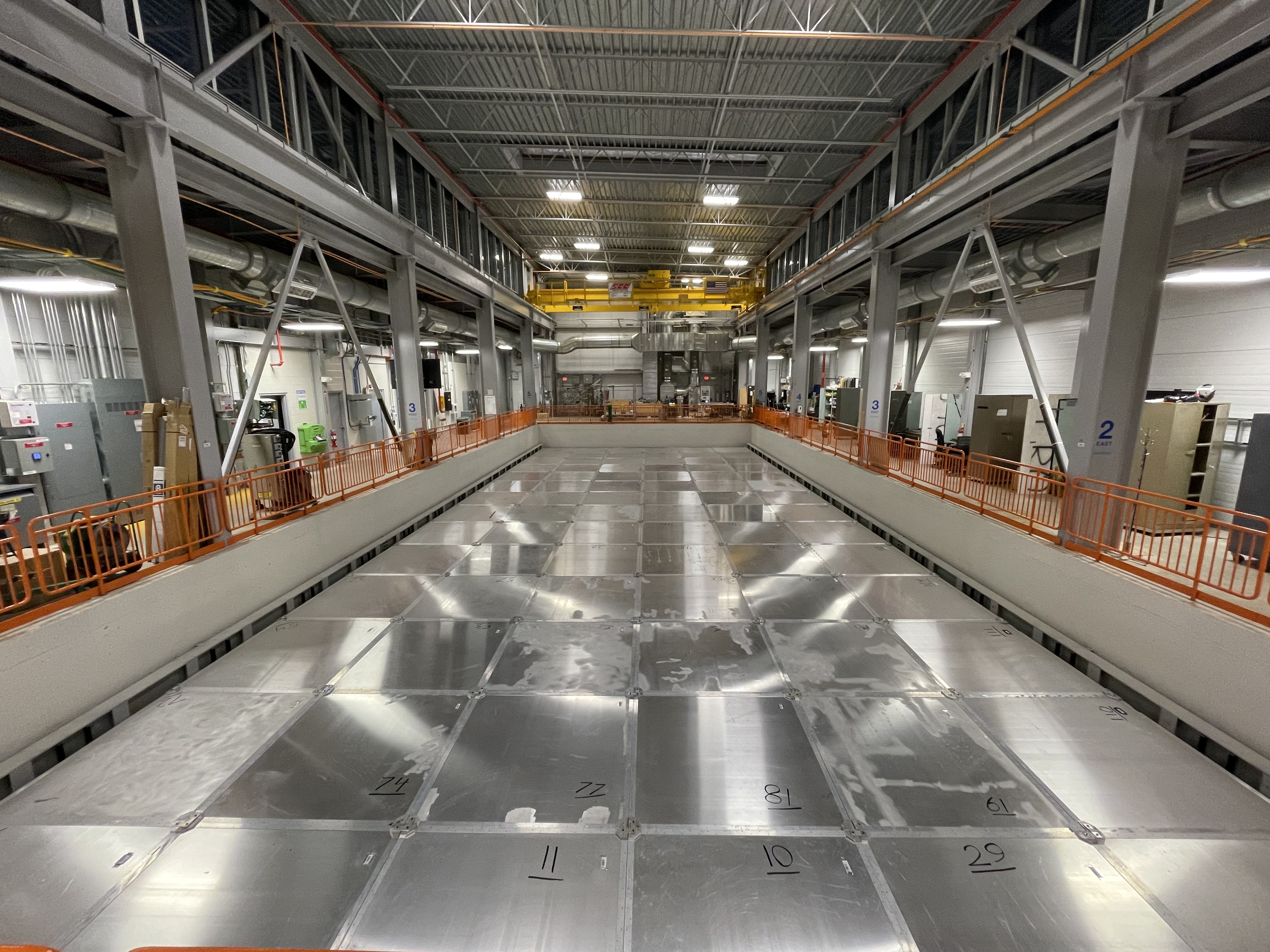}
    \caption{Top CRT horizontal modules installed above the ICARUS detector.}
    \label{fig:TopInstalled}
\end{figure}
\begin{figure}
    \centering
    \includegraphics[angle=270,width=0.40\textwidth]{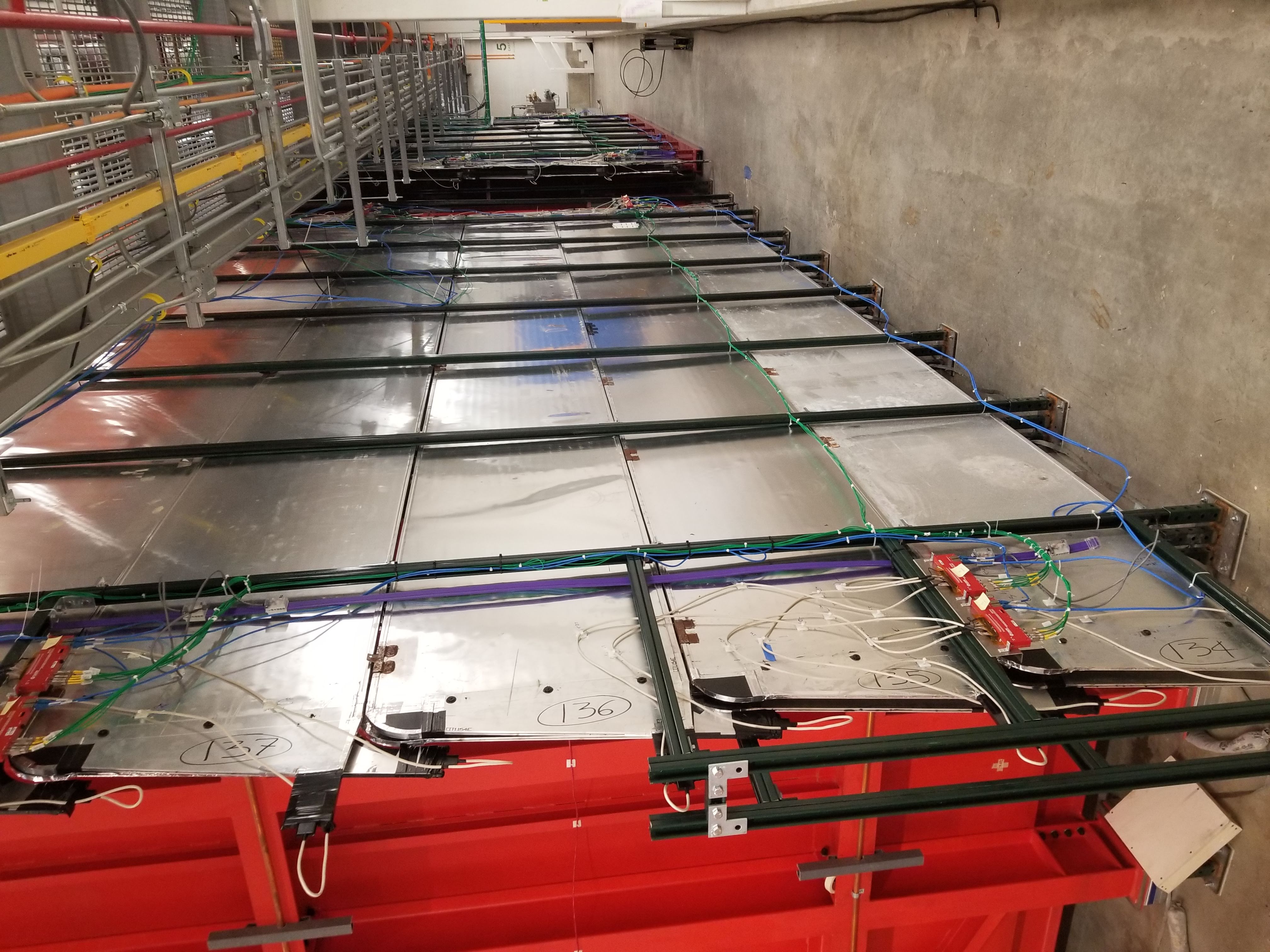}
    \caption{The Side CRT East walls post-installation.} 
    \label{fig:east_walls_complete}
\end{figure}

The installation of Top CRT modules was completed in December 2021. A picture of the installed horizontal modules is shown in Figure \ref{fig:TopInstalled}. The Side CRT modules were installed prior to the commissioning of ICARUS, mostly during 2020 and 2021. A lateral wall of the installed Side CRT is pictured in Figure \ref{fig:east_walls_complete}. The Bottom CRT
was installed in 2017 before the cryostat installation.
\par The Side, Top, and Bottom CRT modules were tested before and after their installation to check the electronic functionality of every channel. Data are transferred to servers via Ethernet cables connected to each module's FEB in daisy chain configuration. For the Side and Top CRT, a voltage of 5.5 V is distributed to each FEB power lines assembled at Fermilab which draw a current of 3.1 and 6.3 A, respectively. Information regarding cable lengths, SiPM bias voltages, module positions, etc. are stored in a SQL database.

In the case of the Bottom CRT, each FEB is powered with 6.3~V and draw a current of 0.32~A. Each FEB is calibrated individually. The average response of the 64 pixels within an FEB is set by the HV value. The individual pixel responses are equalized using cosmic muon data and computing a set of 64 calibration constants. These calibrations constants are stored in a SQL database together with module position, FEB address numbers, trigger threshold and trigger configuration options. 

\par The Top, Side, and Bottom CRT modules operate in self-trigger mode. For the Top and Side CRT modules, whenever an FEB receives signals on SiPM inputs exceeding the threshold or a signal on either T0 or T1 inputs, it initiates the readout of all 32 SiPM channels and stores the data in an internal buffer. Servers periodically poll data from the buffer for all FEBs simultaneously, every 40 ms for the Side CRT FEBs and 80 ms for the Top CRT FEBs. 

For the Bottom CRT the readout scheme is similar: every channel above threshold is readout by the FEB, sent to the USB module, and then to DAQ to form an event. A second trigger scheme is implemented directly into the firmware of the FEB FPGA and reads out only geometrically overlapping channels above threshold.

The ICARUS data acquisition system, based on the \textit{artdaq} software development toolkit \cite{artdaq}, uses a \textit{BoardReader} application to combine Top and Side CRT self-triggers within 10 ms into data fragments. Every time an ICARUS global trigger \cite{icaruspaper} occurs, the DAQ acquires CRT data fragments within $\pm$ 25 ms around the trigger time. The global trigger is generated by an ICARUS trigger system FPGA when a majority of discriminated signals, coming from pairs of PMTs of the LAr scintillation light detection system, is in time coincidence  with the BNB (1.6 $\mu s$) and the NuMI (9.5 $\mu s$) beam spills. The Side CRTs use the same configuration for all the FEBs, namely bias voltage, triggering channels, channel gain, trigger thresholds, while for the Top CRT each channel has his own bias voltage according to its nominal value provided by the producer.
For the Bottom CRT, data from multiple FEBs are combined into 1 data fragment if the events happen within 64~ns. The time for each of the fragment is assigned based on the 32-bits counter coming from each FEB and the USB timestamp. At the beginning of each run, the DAQ loads configuration parameters into the Bottom CRT FEBs. These parameters include trigger thresholds, 64 gain constants, trigger scheme, etc.

%% file: OBinstallation.tex
\par 
After the completion of the Top CRT commissioning, 
the concrete overburden was installed above the Top CRT. It consists of three 1 m thick concrete layers (6 m water equivalent thickness). The first layer, located 10 cm above the Top horizontal CRT, is composed of 0.5 m wide blocks of low radioactive concrete. For the second and third layer, concrete blocks repurposed from previous Fermilab experiments were used. Before their installation, the blocks were tested with a Geiger counter certifying a radioactive level compatible with the environmental one. 



\begin{figure}[t]
    \centering
    \includegraphics[width=0.49\textwidth]{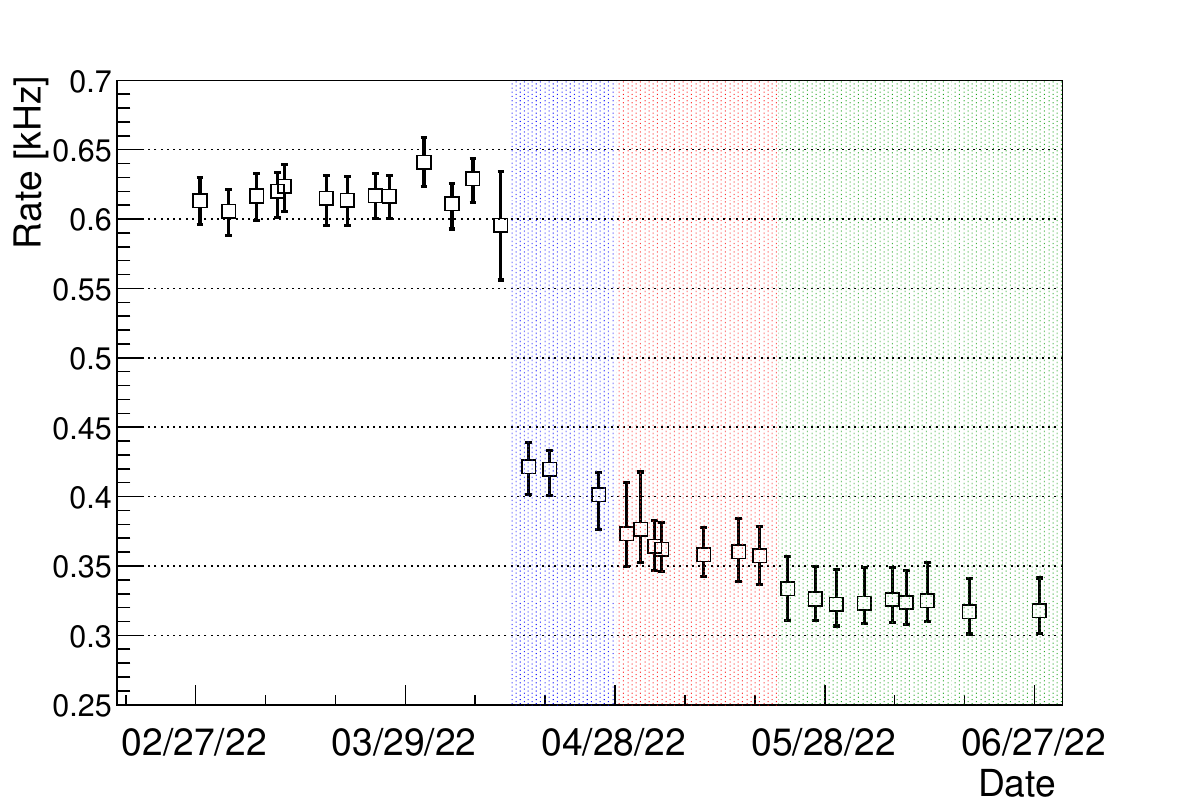}
    \includegraphics[width=0.49\textwidth]{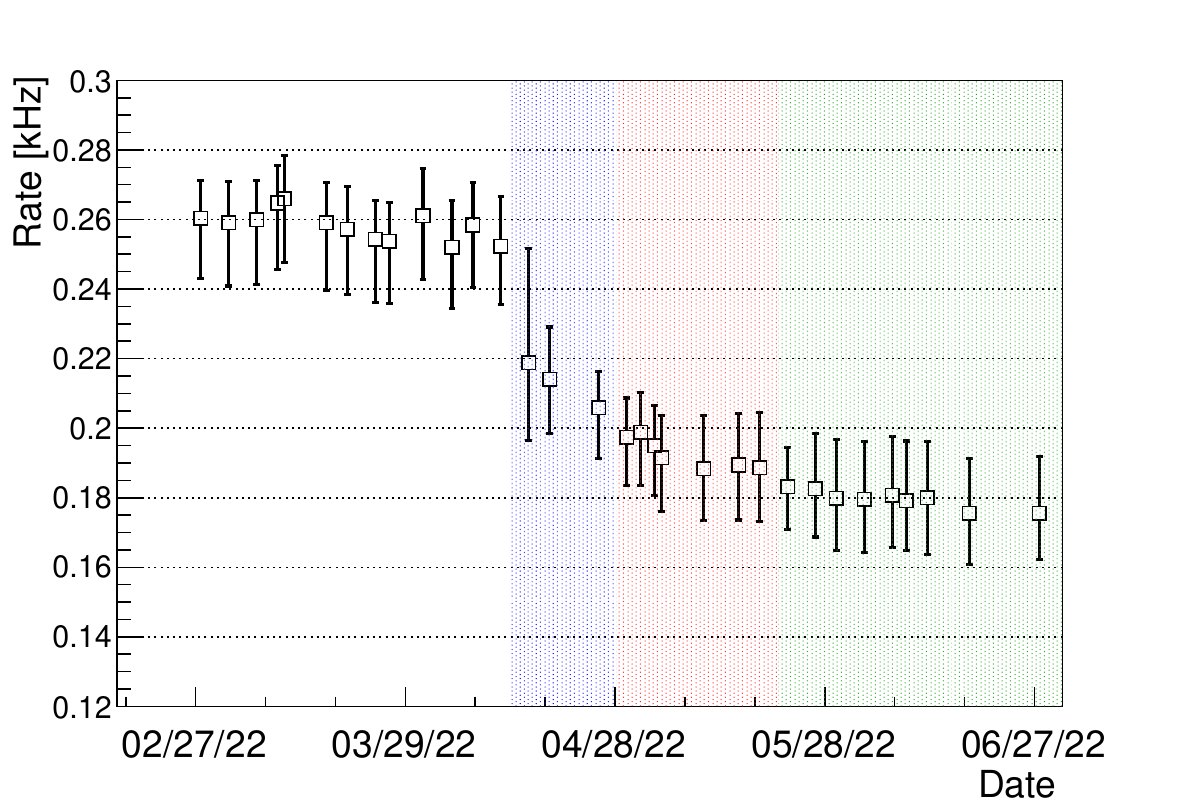}
    \caption{Average trigger rate as a function of time for a subset of Top CRT horizontal (left) and vertical (right) modules. The rate decrease correlates with the progress of the installation of the 1 m overburden layers: first layer (blue), second layer (red) and third layer (green). Error bars represent the standard deviation over the modules.}
    \label{fig:HorRates}
\end{figure}

 The nominal event rate of horizontally installed Top CRT modules before the installation of the overburden was $\sim$~620 Hz, while the rate for vertically installed modules was $\sim$~250 Hz. During the installation of the overburden, day-by-day monitoring of the average cosmic ray rate of each CRT module verified its reduction. In Figure \ref{fig:HorRates} the average cosmic particles rate as a function of time is shown for a sub-sample of horizontal (left) and vertical (right) modules. After the full installation of the overburden the Top CRT cosmic rate per module decreased to $\sim$~330 Hz for the horizontal plane and to $\sim$~180 Hz for the vertical rims, in agreement with the soft electromagnetic component removal mediated by the concrete. The new average rate per module is in agreement with the expected muon-only component of the cosmic spectrum. 
\par With the completion of concrete overburden installation in mid June 2022, the ICARUS detector started to take physics quality data.


%% file: TimingCalibration.tex

\par In order to achieve nanosecond-level timing synchronization between the CRT and other detector subsystems for cosmic muon tagging, a timing calibration is essential. This involved determining delays along synchronization lines from time clock sources to each CRT FEB through an extensive measurement campaign.
\par For each trigger in a CRT module, the FEB generates two distinct timestamps for the event, measured in nanoseconds relative to the acquisition of the T0 and T1 reset signals. These timestamps are then processed offline to associate an absolute timestamp for every CRT interaction which can be compared to events recorded by the other detector subsystems. 
\par
\begin{figure}
    \centering
    \includegraphics[width=0.7\textwidth]{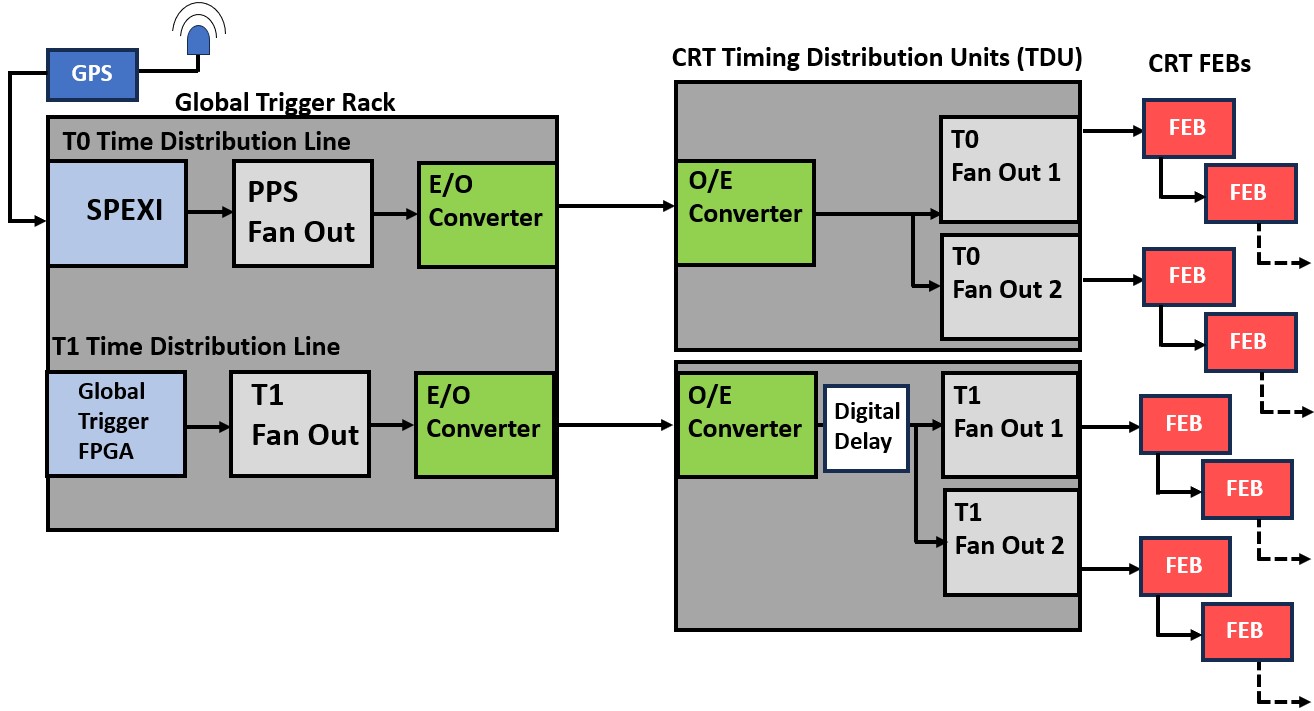}
    \caption{Scheme of the distribution lines of T0 (top line) and T1 (bottom line) reset signals where E/O and O/E refer to Electrical-to-Optical and Optical-to-Electrical, respectively. T0 reset is a PPS signal generated by the SPEXI on the base of a GPS signal. T1 reset is the ICARUS global trigger. 
}
    \label{fig:timinglines}
\end{figure}

 Synchronization signals are distributed to the Side and Top CRT FEBs T0 and T1 LEMO connectors through a Timing Distribution Unit (TDU) as shown in Figure \ref{fig:timinglines}. A PPS signal is sent through RG174 LEMO cables to the FEBs internal counter T0 which resets every second and is used as an absolute time reference. The PPS is an absolute GPS timing signal generated by a SPEXI board \cite{icaruspaper} sent to a Fan Out unit and distributed to the three different CRT subsystems. There are three TDU racks for the CRT subsystems: a Top TDU that distributes timing signals to all top FEBs, an East TDU that distributes signals to the Side-East and Side-South FEBs, and a West TDU that distributes signals to the Side-North, Side-West and Bottom CRT boards. From the CRT TDU, the PPS signal is propagated to timing distribution lines consisting of a bus of lemo cables connected via a Y connector plugged into each of the T0 ports at every step of the chain.
\par In the same way, the T1 counter receives a reset signal from the ICARUS global trigger providing a time reference for the synchronization of the PMT and CRT signals. The T1 timing distribution line is similar to the T0 one but in addition a digital delay unit is used inside the TDU as shown in Figure\ref{fig:timinglines}. As mentioned in Section \ref{sec:TopCRT}, the acquisition of T1 reset events induces a 22 $\mu$s dead time; to avoid T1 reset occurring during the TPC drift readout window the T1 signal is delayed by 2~ms.  
\par
\begin{figure}
    \centering
    \includegraphics[width=0.7\textwidth]{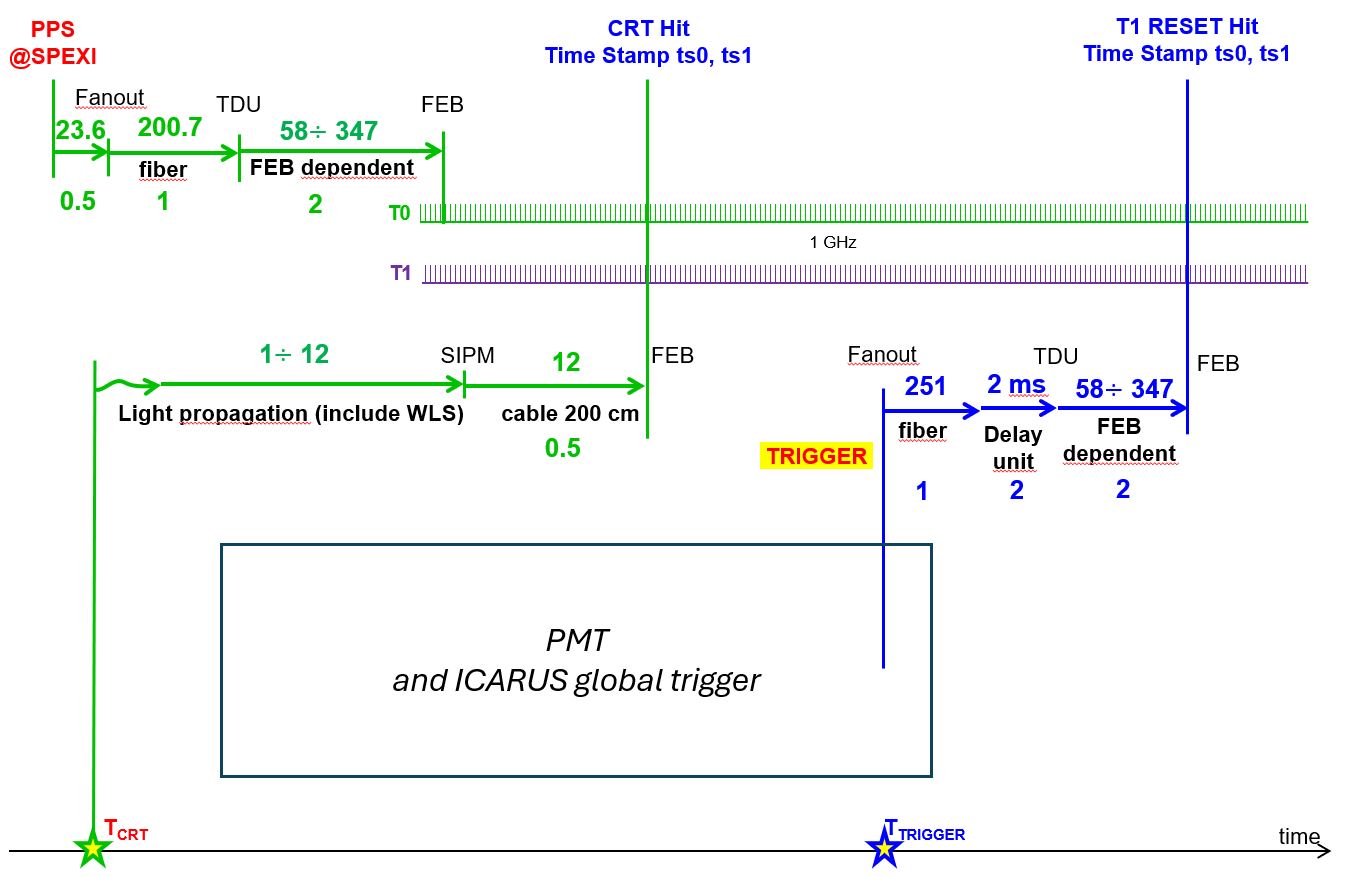}
    \caption{ Scheme of Top CRT signal propagation time (in ns): numbers above the arrows indicate the measured/estimated delays while numbers below are the corresponding systematic error.  The two stars on the time axis correspond to the time at which the cosmic ray energy deposit produces light in the Top CRT (green) and the time at which a global trigger signal is generated from the ICARUS trigger system (blue).}
    \label{fig:propagation-time}
\end{figure}

 The delay contributions used to derive the absolute hit time for the Top CRT are shown in Figure \ref{fig:propagation-time}.

A similar procedure was applied to determine the delay contributions for each Side CRT FEB. As the timing distribution chain is identical up to the CRT TDUs, the same value of 23.6 ns is adopted for the signal propagation delay from the SPEXI to the Fanout within the Global Trigger Rack. The delay between that fanout and the East/West TDUs was similarly measured using the two-ways technique: delays for the East TDU are 82.7 ns and 106.73 ns while the delays for West TDU are 201.4 ns and 225.71 ns for the T0 and T1 lines, respectively, with a 1 ns systematic error on all values accounting for distribution width and electronics contributions. The signal propagation delays from the East/West TDU rack to each FEB were calculated using known cable lengths and a precisely measured signal propagation speed.

\par For the Bottom CRT the timing calibration was finalized using coincidence cosmic events recorded in coincidence with the top and Side CRT hits. Individual cable lengths were measured during and after the installation of the modules, those values together with the delay of the processing and digitization time of the FEB are stored in an SQL database and applied during the data taking and event building. The delays are now with a precision of about 1 clock cycle (16~ns).



%% file: DetectorResponseCalibration.tex
\begin{figure}
    \centering
    \includegraphics[width=0.70\textwidth]{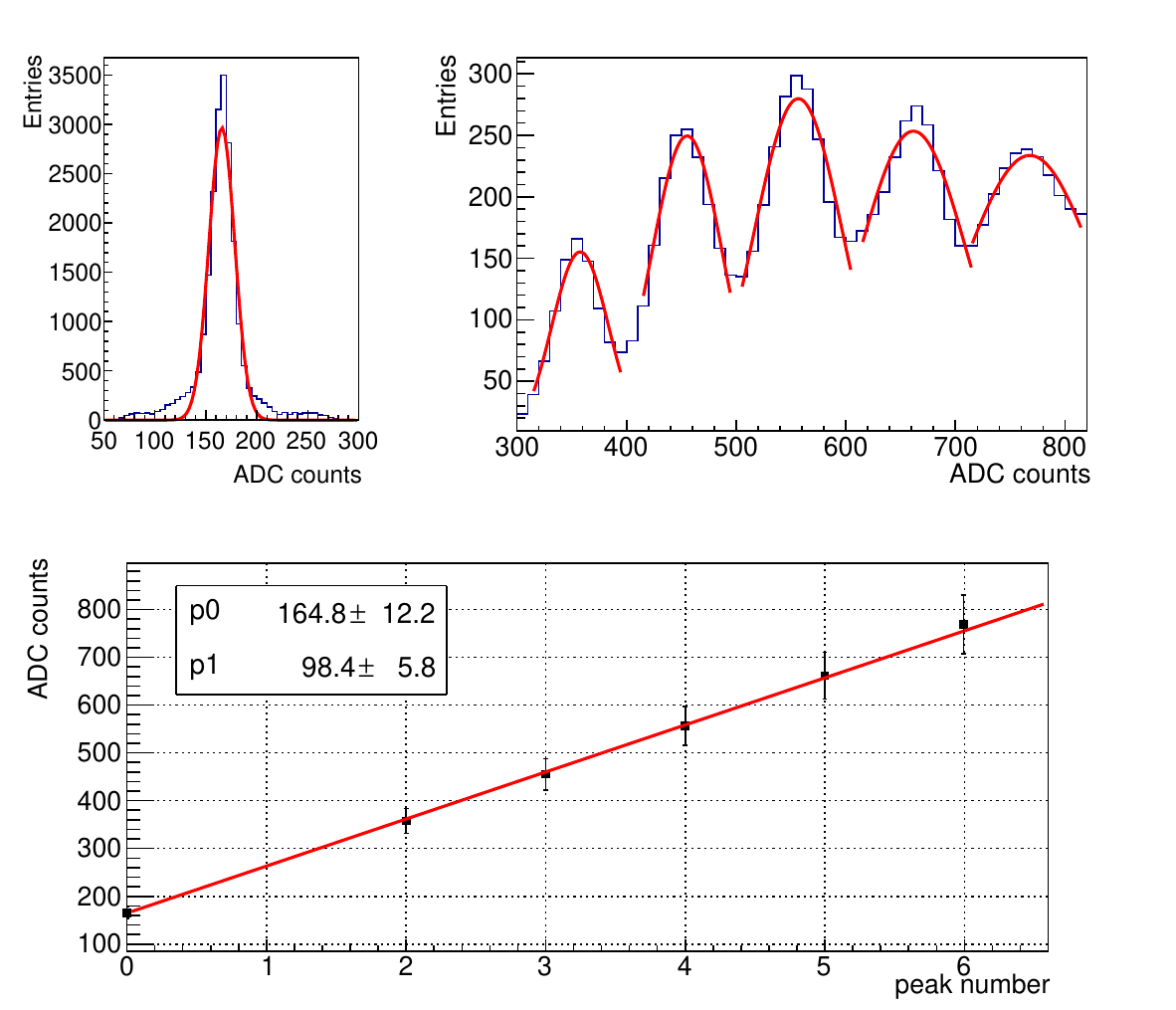}
    \caption{Calibration analysis of one Top CRT channel. Top left: region of the ADC spectrum showing the pedestal distribution. Top right: region of the ADC spectrum of one CRT channel with the Gaussian fit superimposed over each peak. Bottom: mean ADC values obtained from the fit of each peak as a function of $n_{p.e.}$. The error bars of each point represent the standard deviation of each peak obtained from the Gaussian fit. The parameters p0 and p1 are the y-intercept and slope of the linear fit, respectively.}
    \label{fig:calib_ADC_vs_peaknum}
\end{figure}

A calibration procedure was developed to estimate the pedestal and gain values of each FEB channel in order to monitor and equalize the detector response across the CRT modules. Gain and pedestal for all channels of each FEB are determined by analyzing the corresponding ADC charge spectrum.

The gain G was obtained as the distance between the p.e. peaks in the charge spectrum, knowing that:
\begin{equation}
    G = \frac{ADC - ped}{n_{p.e.}}
\end{equation}
where $n_{p.e.}$ is the number of quantized photoelectrons detected by the SiPM, \textit{ADC} is the mean value of the corresponding peak, \textit{ped} is the mean value of the pedestal for the analyzed channel and \textit{G} its gain. The charge spectrum is processed with an iterative algorithm that searches for peaks by fitting them with Gaussian distributions, as shown in Figure \ref{fig:calib_ADC_vs_peaknum}, top right. The pedestal is determined by slightly different procedures for the Side and Top CRTs. For the former, the pedestal is obtained from the same ADC charge spectrum used to determine the gain, looking only at the first peak found in the 1$\div$300 ADC range. For the Top CRT, a different ADC spectrum is obtained by triggering the FEB with an external PPS signal, thus selecting only ADC values without the presence of p.e. peaks. In both cases, the pedestal distribution is fitted to a Gaussian distribution (Figure \ref{fig:calib_ADC_vs_peaknum}, top left). The mean ADC value of the pedestal and the fitted p.e. peaks are plotted as a function of their corresponding p.e. number, with the pedestal corresponding to zero p.e., and the gain is obtained as the slope of a linear fit (see Figure \ref{fig:calib_ADC_vs_peaknum}, bottom). For the entire Top CRT the gain constants vary by 7\% among the different channels while the for the entire Side CRT the corresponding variation is 4\%.
\par For the Bottom CRT the calibration and equalization of the response of individual channels was performed using cosmic muons. A first calibration of individual HV channels was performed comparing the average response of each individual MAPMT. A detailed calibration of individual pixels was then carried out for each MAPMT pixel using both single p.e. hits and overlapping muon hits in both layers of the CRT modules. The amplification of individual channels  was adjusted using the FEB's capability, utilizing the overlapping muon hits for calibration. This resulted in 576 (9 modules with 64 individual strips) gain constants. The difference between the response of individual channels of the bottom CRT is less than 7\%.

%% file: CRTHitReconstruction.tex
\begin{figure}
    \centering
    \includegraphics[width=0.7\textwidth]{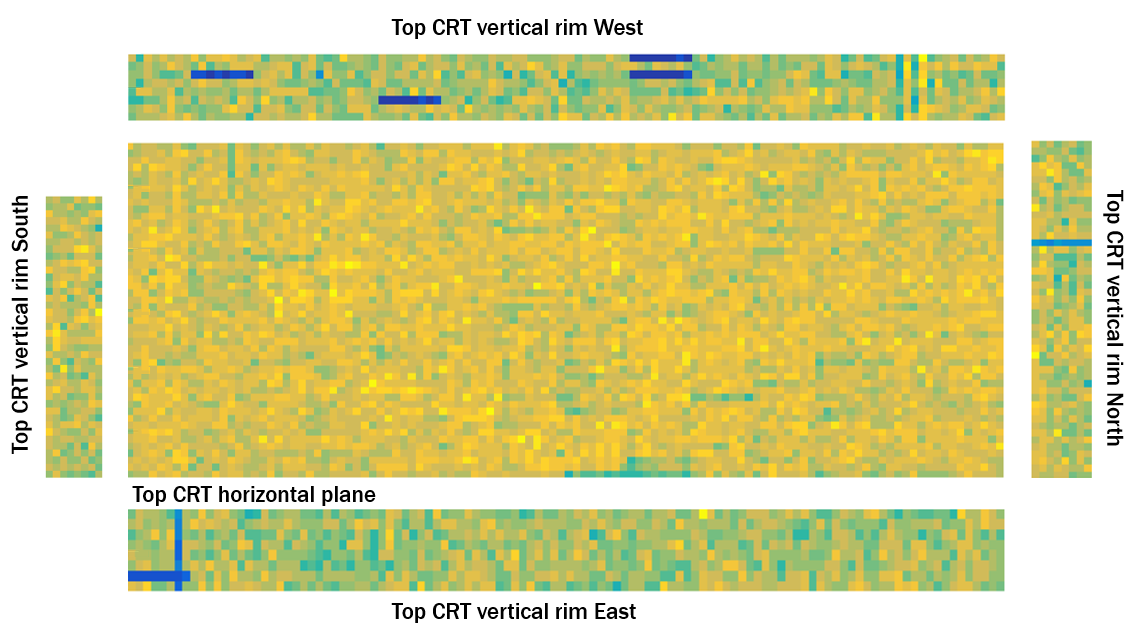}
    \caption{Distribution of the CRT hits reconstructed in the different regions of the Top CRT. Blue regions correspond to malfunctioning channels.}
    \label{fig:TopCRTHitReco}
\end{figure}

The first step of the hit reconstruction is the extraction from raw data of FEB ID number,
 timestamps of the T0 and T1 counters, ADC values of all 32 SiPM channels. The CRT hit position relative to the center of the CRT module is determined from the triggering channels, using a different procedure for Top and Side CRTs. For the latter, only channels with signal amplitude above a threshold of 7.5 p.e. are selected. For the Top CRT, the hit is provided by quadruple coincidence (the four SiPMs of the two module layers) required by the internal triggering logic. The local hit position is then translated into the global ICARUS reference system. The CRT hit timestamps are corrected for the light propagation along the fibers. The distribution of reconstructed hits for the Top CRT of an ICARUS run acquired during the commissioning is shown in \figurename{ \ref{fig:TopCRTHitReco}}. The granularity of the distribution is determined by the 23$\times$23 cm$^2$ size of the overlapping region of two orthogonal strips. Except for inefficiencies due to few broken SiPM channels (blue bars) the distribution shows a uniform response of the Top CRT detector.

\par For the Side CRT the coincidence of adjacent layers is reconstructed at software level. Hit scintillator strips are identified by selecting the channel within each FEB that generated the trigger signal, using the highest charge amplitude as a discriminator. 
If the scintillator strip is read out at both ends by two different FEBs above threshold with hits within 150 ns, the longitudinal position can be reconstructed using the FEB timestamps with: 
\begin{equation}\label{eq:zpos}
   z=(T_B-T_A)/2\cdot V
\end{equation}
where $T_A$ and $T_B$ are the timestamps recorded by the FEBs at the two ends of the strip, $V$ is the speed of light along the WLS fiber, and $z$ is the reconstructed CRT hit position.
\par The reconstruction of the longitudinal $Z$ position is performed on both the inner and the outer layers. The CRT hit position is given by the average of the two reconstructed longitudinal positions.
In the case of the South wall, which exploits an X-Y configuration and the outer layer is read out on one end only, the CRT hit position along the horizontal East-West direction is provided by the position of the vertical strip hit in the outer layer, while the vertical coordinate is given by the position of the strip hit in the inner layer. In the case of the downstream North wall (which uses cut parallel scintillator strips read out on just one end), the CRT hit position is obtained as the average position of the strips hit in the two layers.

The Bottom CRT hit position is reconstructed using the readout on one end of the module only. The Bottom CRT hit position is provided by the position of the strip hit in the module, this only provides the position of the hit with respect to the position of the module. Transversal position is not known, as the modules are read out only on one end.

%% file: timeResolution.tex
The time resolution of the FEBs was determined by evaluating the time difference between consecutive reference PPS signals distributed to the various FEBs. On average, the measured standard deviation of the boards is $\sim$2.4 ns.
\par
\begin{figure}
    \centering
    \includegraphics[width=.7\textwidth]{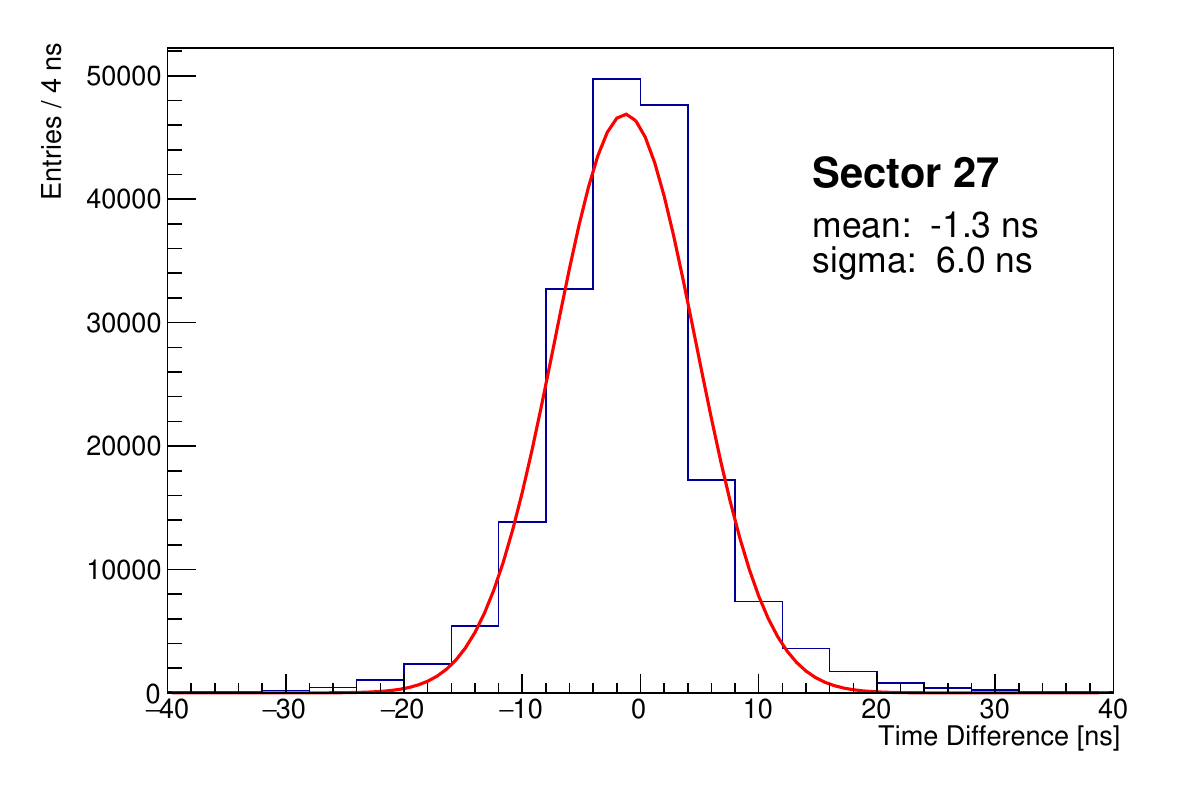}
    \caption{Time difference between upper and lower Top CRT External hodoscope modules for CRT Hits in one of the coincidence sectors.}
    \label{fig:TopCRTTimeResolution}
\end{figure}

 The time resolution of the Top CRT modules was evaluated using the external hodoscope described in Section \ref{sec:TopCRT}.
The measurement was conducted using cosmic particles crossing the hodoscope almost vertically by selecting CRT hits in the same triggering sector on the top and bottom hodoscope modules. A distribution of the time differences between hits in the top and in the bottom telescope modules was obtained for each of the coincidence sector. An example for one of the 64 coincidence sectors is shown in Figure \ref{fig:TopCRTTimeResolution}, with, superimposed, a Gaussian fit distribution. The distribution has a standard deviation of $\sigma_{{\Delta}T}\sim$ 6 ns. 
Assuming that both the top and bottom modules of each sector are characterized by the same time resolution, the intrinsic module time resolution is $\sigma_{Top CRT}=\sigma_{{\Delta}T}/\sqrt{2}$. This quantity was evaluated for all the 64 hodoscope coincidence sectors, giving an average value $\sigma_{Top CRT}<4.4$ ns.
This upper limit for the Top CRT time resolution is expected considering the intrinsic $\sim$ 2.4 ns resolution of the FEBs and the contribution of up to few ns of the light propagation in the plastic scintillator and in the WLS fiber.
The time resolution is assumed to be the same for all the Top CRT modules.

\begin{figure}
    \centering
    \includegraphics[width=.6\textwidth]{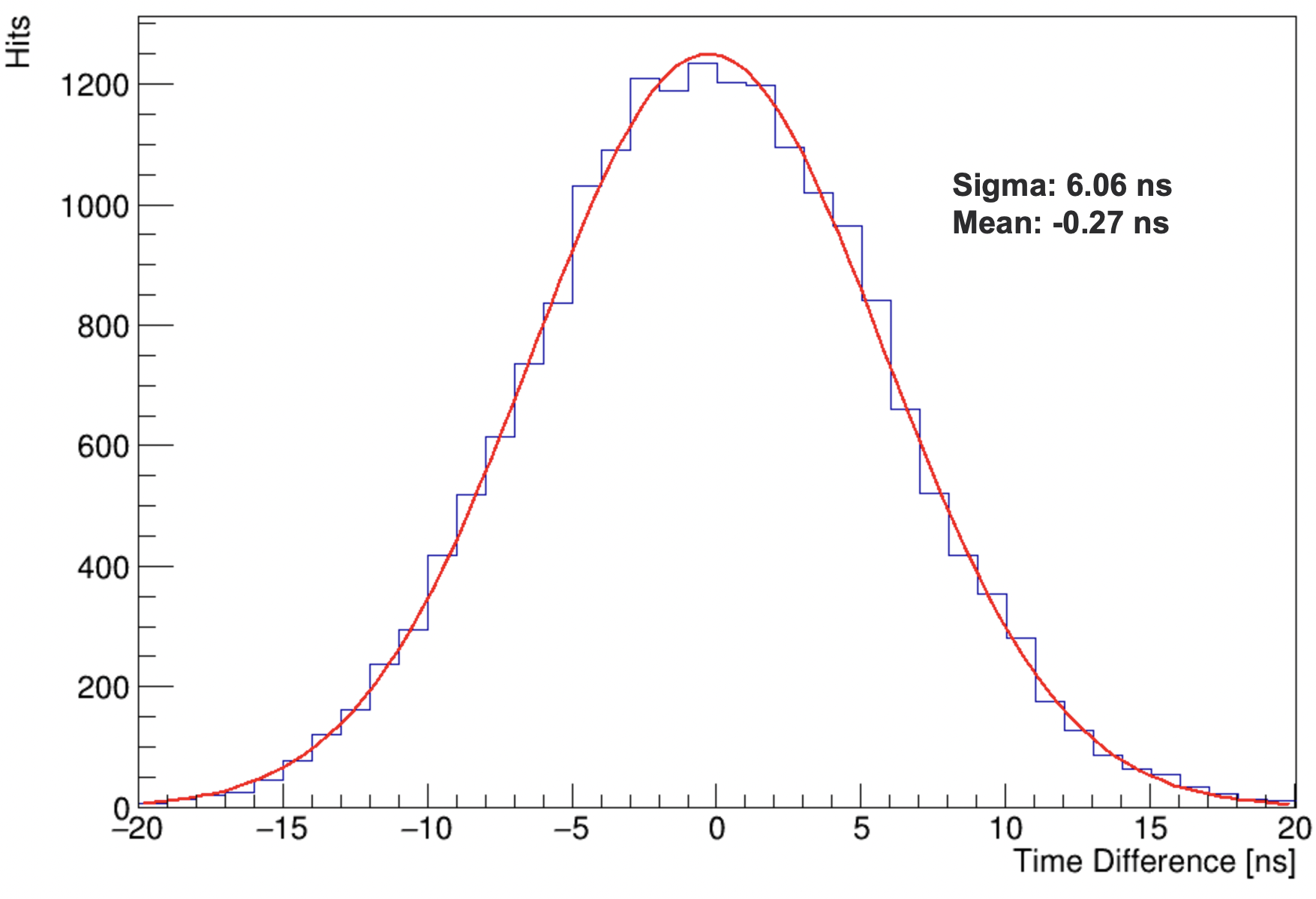}
    \caption{Time difference between the hits from East-Center wall and East-North walls. }
    \label{fig:SideCRTTimeResolution}
\end{figure}

The Side  CRT modules time resolution was evaluated using two overlapping CRT walls, the East-Center wall and East-North wall from Figure \ref{fig:sideCRTgeometry}. The  East-Center wall was shifted north by 3.35 m to have an overlap of $\sim$2 m with the East-North wall and data were collected with this configuration. The distribution of the time differences between the hits from the East-Center and East-North walls is shown in Figure \ref{fig:SideCRTTimeResolution}. The mean value shows an offset of ~1 ns and  $\sigma_{{\Delta}T}\sim$ 6.06 ns similar to the time resolution of the Top CRT.

The time resolution of the Bottom CRT was measured using two overlapping CRT modules at the ProtoDUNE\cite{BottonCRTResolution} detector which uses the same modules from Double Chooz. The  time resolution was measured to be better than 60 ns.

%% file: efficiency.tex
The efficiency of the Side CRT strips was evaluated using a test stand consisting in a pair of MINOS modules and two 3 cm x 2 cm scintillator tiles coupled to a PMT used as an external trigger. The detection efficiency for through-going muons was determined at different positions in the MINOS module. At the middle of the module (4 m), the efficiency for two layers in coincidence was found to be 98\% and at 6 m it was 95\%.
\par A test stand was set up to measure the efficiencies of the Top CRT modules at the INFN-LNF production facility. Six modules were stacked on top of each other spaced 30 cm apart. The efficiency of the inner modules was evaluated using the logical AND of the signals from the top and bottom modules as a reference.
\par The expected efficiency of the bottom CRT modules in the ICARUS configuration with no overlaps is above 95\%. This was measured at the University of Chicago and then confirmed in the Double Chooz experiment and it is related to the internal efficiency of a module. Each module is made of a staggered 2-layers of scintillating strips, so in the trigger scheme implemented in ICARUS in which we readout only geometrically overlapping strips the two half of the end strips at the edge of each modules induce a small geometrical inefficiency. We couldn't test the efficiency for the Bottom CRT in ICARUS, we have only modules oriented in the X or Y direction and they are single modules and not overlapping X and Y as in Double Chooz.

%% file: tagging.tex
\par
\begin{figure}
    \centering
    \includegraphics[width=.7\textwidth]{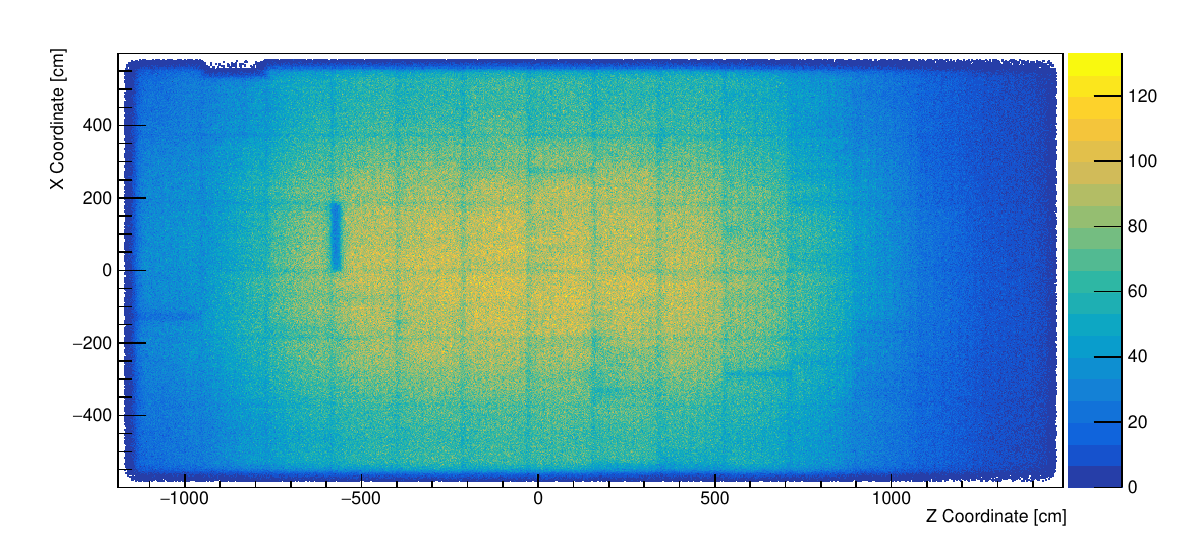}
    \caption{The distribution of X-Z positions from data for TPC tracks projected into the Top CRT horizontal plane for tracks with a CRT hit within 40 cm of the extrapolation point.}
    \label{fig:ProjectionMap}
\end{figure}

 A measurement of the cosmic ray tagging efficiency has been performed for the fully commissioned CRT by exploiting an algorithm that matches TPC tracks with CRT hits. In this paper we report on the Top system, which provides coverage for approximately 77\% of the cosmic ray flux incident on the ICARUS TPCs. Data were collected during a period when there was no beam using the PMT system
as a trigger. For tracks that were well-reconstructed by the TPC, a straight-line projection was used to calculate the intersection of the track with the CRT in the absence of multiple Coulomb scattering. A Monte Carlo study of truth matches between TPC tracks and CRT hits indicates that 97.6\% of those tracks would have a distance of the extrapoled track position on the CRT plane and the matched CRT hit smaller then 100 cm. Using a sample of approximately 4.1 million reconstructed tracks, the tagging efficiency for the Top CRT horizontal plane was found to be 94.9\%. In Figure \ref{fig:ProjectionMap} shows the distribution of the Z-X positions from data of the TPC track extrapolation on the Top CRT horizontal plane for tracks with a CRT hit within 40 cm of the projection point. The 40 cm cut for the Top CRT results in high purity ($>$97$\%$) when associating TPC tracks with Top CRT hits.

%% file: Conclusions.tex
ICARUS is located at surface level, where cosmic ray background is significant. To mitigate this background, the experiment employs a Cosmic Ray Tagger that surrounds the cryostat, consisting of extruded organic scintillator equipped with wavelength-shifting fibers and silicon photomultipliers. The installation was finalized in 2021,followed by commissioning in 2022. The average cosmic ray rates were halved thanks to the six-meter concrete overburden placed above the top CRT. Since commissioning, the CRT system has operated smoothly and has been actively collecting physics data since 2022.

To achieve precise nanosecond-level timing synchronization between the CRT and other detector subsystems, an extensive calibration process was carried out. This involved measuring delays from time clock sources to the CRT FEBs and exploiting special signal recorded by the FEBs when receiving the external PPS clock or the detector trigger signal.   Calibration also required measuring signal propagation delays across various CRT subsystems, the delays were accurately quantified. Detector response calibration was implemented to equalize the gain and pedestal values across FEB channels, using ADC charge spectra. This calibration is used in the CRT hit reconstruction.

Time resolution was evaluated by measuring differences between consecutive PPS signals  distributed to the various FEBs, resulting in average standard deviation of about 2.4 ns. The time resolution of the top CRT system is less than 4.4 ns and together with the accurate timing delays measurements enables a separation between cosmic ray tracks and those originating from inside the detector. Since the Top CRT is located several meters away from the liquid argon active volume, the average time of flight (~25 ns, due to the long flight lengths) of tracks between CRT and TPCs is long enough to distinguish the particle origin (outside or inside the TPCs). In addition, the tagging efficiency between TPC tracks and top CRT hits is reported to be better than (97$\%$).
\par A triple match among the CRT signals, optical flashes from the PMT and tracks from the TPC systems can be used to further reduce the cosmic background. In addition a match between CRT and  PMT  timing signals can be exploited to reject cosmic ray background; this will be the object of an upcoming paper.
